\documentclass[12pt,letterpaper,twoside]{report}
\usepackage{amsfonts} % para incluir fonts y
\usepackage{amssymb}  % simbolos de la AMS
\usepackage{graphicx} % standard LaTeX graphics tool
\usepackage{amsmath}  % for including eps-figure files
\usepackage{amsmath}  % paquete matemático
\usepackage[latin1]{inputenc}
\usepackage{latexsym}
\usepackage[amssymb]{SIunits} %para las unidades del SI
\usepackage{color}
\newcommand{\M}{{\cal {M}}}
\newcommand{\B}{{\cal {B}}}

\newcommand{\HH}{{\cal {H}}}
\begin{document}
\pagestyle{empty}
\begin{titlepage}
\begin{center}
{\large \bf UNIVERSIDAD DE LA HABANA.}

\vspace{0.3cm}

{\large \bf FACULTAD DE FÍSICA.}

\vspace{0.3cm}

{\large \bf ICIMAF.}

\begin{figure}[h!]
\begin{center}
\includegraphics[height=4cm,width=4cm,angle=0]{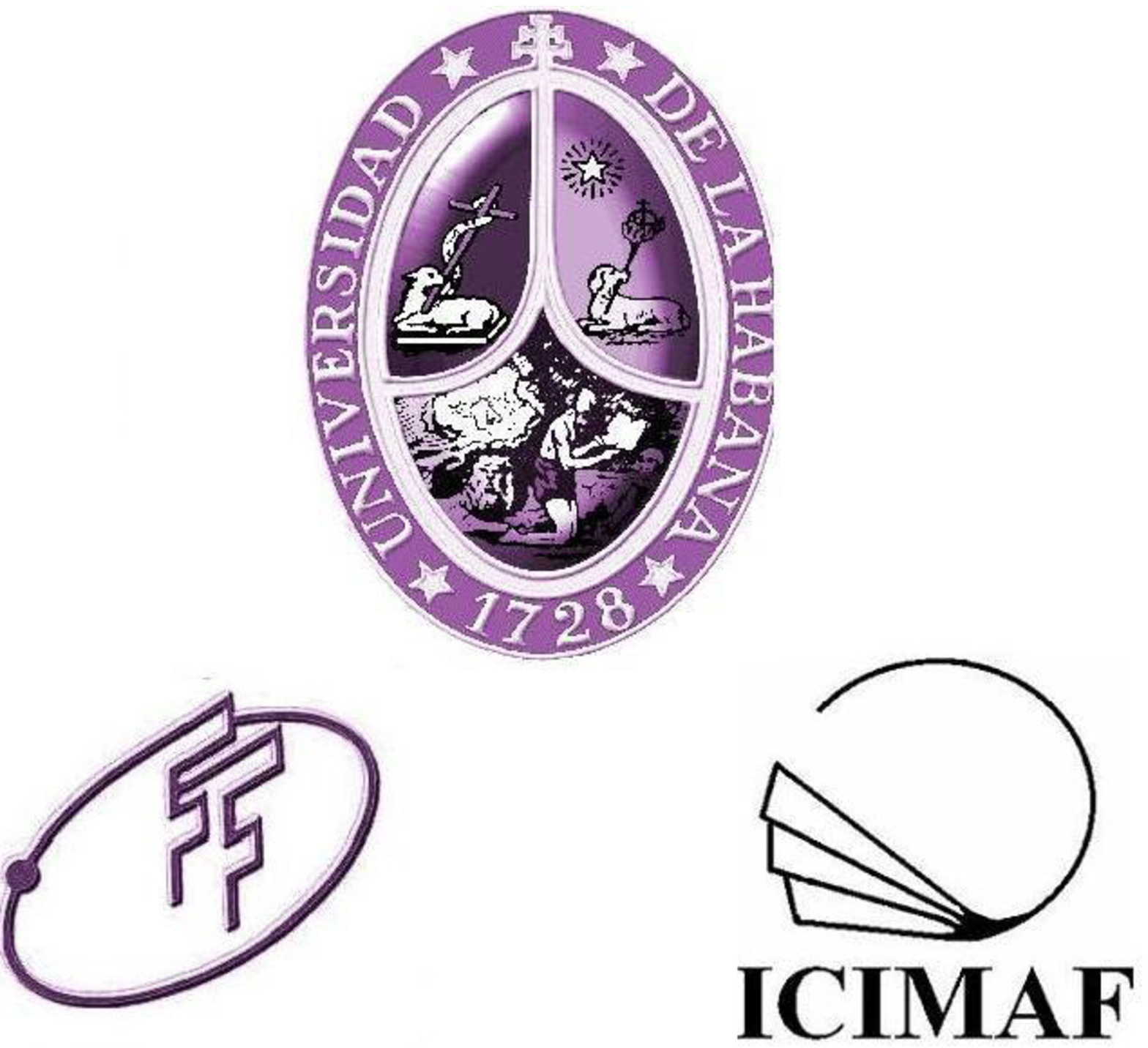}
\end{center}
\end{figure}
\vspace{0.1cm}

{\Huge \bf Dinámica de una fuente \\
magnetizada de neutrones \\
autogravitante. }

\vspace{0.4in}

{\large \bf TESIS DE DIPLOMA}\\

%\vspace{0.8cm}

%{\large  \it presentada en opci\'on al grado de} \\
%{\Large  \bf Master en Ciencias F\'{\i}sicas}\\
\end{center}

\vspace{0.6in}

{\large \bf Autor:}\hspace{0.3cm}{\large Daryel Manreza Paret.} \\

{\large \bf Tutores:} \hspace{0.01cm}{Dra. Aurora Pérez Martínez,
ICIMAF, Cuba.}

{\large }\hspace{2.05cm} {MSc. Alain Ulacia Rey, ICIMAF, Cuba.}
\vspace{0.5in}
\begin{center}
{\large  Ciudad de La Habana}\\
\vspace{0.5cm}
{\large  2008}\\
\end{center}
\end{titlepage}

%%%%%%%%%%%%%%%%%%%%%%%%%%%%%%%%%%%%%%%%%%%%%%%%%%%%%%%%%%%%%%%
%%%% Dedicatoria
\newpage\voffset=2in \thispagestyle{empty} \hoffset=4.5in
\noindent {\Huge \it \ \ \ \small{para mi abuela y mi mamá}}

%%%% Agradecimientos
\newpage
\hoffset=0in \voffset=1in \vspace{3in} \thispagestyle{empty} {\large
\bf    Agradecimientos}

\vspace{0.5in}

{\it
 Llegar al final ha implicado la suma de las colaboraciones de
muchos en sus respectivas trincheras, sin los cuales el producto no
hubiese sido tan preciado.

Primeramente mi mamá, que me compromete con su amor y su tenacidad a
hacer las cosas con entrega y esmero, además de estar ahí siempre,
al alcance de mis manos. A mi abuela, mis tías, mis hermanos y mis
primos, esa gran familia que hace todo posible, me animan y
depositan en mí su confianza, a la que hago ahora gala. A mi papá,
por tanta ayuda, y porque a pesar de la distancia, se que se siente
muy orgulloso por cada logro. A Karel y Yuliet, los amigos de
siempre.

A Aurorita, por guiarme en los estudios e iniciarme en la
comprensión de la física de los objetos compactos, por compartir
conmigo durante estos años y constituir un ejemplo como
investigadora y como persona. A Alain, que con sus enseñanzas sobre
gravitación me motiva a investigar esta materia, y quien despierta
curiosidades merece más que estas gracias. A Alejandro Cabo, Zochil,
Augusto, Elizabeth, Alain Delgado, Nana y Eru.

A mi novia Anabel, depositaria también de mi tiempo, quien lo
diversifica y me devolvió al trabajo con ideas frescas. A su
familia, que es ya mía.

A los profesores de la Facultad de Física, de quienes hemos
aprendido el rigor científico y la ética profesional. A los amigos
de la carrera, con quienes he compartido cinco años entre estudios y
diversiones.

A todos los que me han ayudado y me han apoyado en la carrera y en
mi vida.

Finalmente gracias...por todo.}

%%%%%%%%%%%%%%%%%%%%%%%%%%%%%%%%%%%%%%%%%%%%%%%%%%%%%%%%%%%%%%%%%%%%%%%%

%%%Resumen
\newpage
\hoffset=0in \voffset=1in \vspace{3in} \thispagestyle{empty} {\large
\bf    Resumen}

\vspace{0.3in}

La din\'amica de un gas autogravitante de neutrones en presencia de
un campo magn\'etico, es estudiada tomando como ecuaci\'on de estado
del gas de neutrones la obtenida en un estudio previo
\cite{Aurora1}. Se trabaja con un espacio Bianchi~I caracterizado
por una m\'etrica tipo Kasner que permite tomar en cuenta la
anisotrop\'ia que introduce el campo magn\'etico. El sistema de
ecuaciones de campo de Einstein Maxwell para este gas se convierte
en un sistema de ecuaciones din\'amicas en un espacio de fase en 4D.
Se obtienen soluciones num\'ericas de dicho sistema. En particular
se encuentra una soluci\'on singular tipo punto para diferentes
condiciones iniciales. F\'isicamente esta soluci\'on singular
podr\'a asociarse al colapso de un volumen local de materia
neutr\'onica en el interior de una Estrella de Neutrones.

\hoffset=0in \voffset=1in \vspace{0.5in} \thispagestyle{empty}
{\large \bf Abstract}

\vspace{0.3in}

The dynamics of a self-gravitating neutron gas  in  presence of a
magnetic field is being studied taking the equation of state of a
magnetized neutron gas obtained in a previous study \cite{Aurora1}.
We work in a  Bianchi I spacetime  characterized by a  Kasner
metric, this metric  allow us to take into account the anisotropy
that introduces the magnetic field. The set of Einstein-Maxwell
field equations for this gas becomes a dynamical system in a
4-dimensional phase space. We get numerical solutions of the system.
In particular there is a unique point like solution for different
initial conditions. Physically this singular solution may be
associated with the collapse of a local volume of neutron material
within a neutron star.
%%%%%%%%%%%%%%%%%%%%%%%%%%%%%%%%%%%%%%%%%%%%%%%%%%%%%%%%%%%%%%%%

%%%%%%%%%%%%%%%%%%%%%%%%%%%%%%%%%%%%%%%%%%%%%%%%%%%%%%%%%%%%%%%%

\voffset=0in % Margen superior de las páginas 1in por default+voffset

\pagestyle{plain}

\pagenumbering{roman} \setcounter{page}{0}

\tableofcontents \pagebreak \pagenumbering{arabic}
\setcounter{page}{1}

%%%%%%%%%%%%%%%%%%%%%%%%%%%%%%%%%%%%%%%%%%%%%%%%%%%%%%%%%%%%%%%%%%%
%Introducción
%%%%%%%%%%%%%%%%%%%%%%%%%%%%%%%%%%%%%%%%%%%%%%
%              Introducción                  %
%%%%%%%%%%%%%%%%%%%%%%%%%%%%%%%%%%%%%%%%%%%%%%
%\chapter{Introducción}
\section{Introducción}

Entre los objetos que habitan en el universo uno de los más
interesantes y que poseen mayor riqueza de fenómenos físicos son las
Estrellas de Neutrones (ENs). El estudio de este tipo de objetos
presenta una gran variedad de atractivos para los físicos, tanto
teóricos como experimentales.

En las ENs concurren una amplia gama de condiciones físicas extremas
como son: altas densidades, elevados campos magnéticos y el hecho de
ser una de las cinco configuraciones estelares (Enanas Blancas,
Estrellas de Neutrones, Huecos Negros, Estrellas Supermasivas y
grupos de estrellas relativistas) en las cuales es preciso tener en
cuenta los efectos de la Relatividad General (RG) \cite{MTW}. Su
complejidad hace que además de la Gravitación haya que tener en
cuenta las otras tres interacciones fundamentales (fuerte, débil y
electromagnética).

En el año 1934  se predijo por primera vez, basándose en cálculos
teóricos, la existencia de una ENs y a partir de esa fecha se
estudiaron muchos modelos para describirlas, pero no es hasta el año
1967 en que se verifica observacionalmente por azar  la existencia
de un objeto con las características de una ENs.

Las ENs no se habían observado antes debido a que no emiten en el
visible, por lo cual fue necesario que pasaran más de 30 años y que
la tecnología permitiera la construcción de radiotelescopios capaces
de detectar las primeras ENs. El descubrimiento de la primera ENs
provocó un aumento en el número de trabajos teóricos encaminados a
describir los procesos físicos que ocurren en el interior de estos
objetos.

Desde hace veinte años los estudios de las Estrellas de Neutrones se
nutren de cuantiosas observaciones que se obtienen gracias a
observatorios que detectan emisiones de rayos X y radiación gamma
colocados en satélites.  Entre los observatorios más conocidos están
el Telescopio Espacial Hubble, el Observatorio de Rayos X Chandra y
el Observatorio de Rayos Gamma Compton.

Actualmente las principales interrogantes concernientes a la física
de las ENs son:

\begin{itemize}
  \item Determinación de cotas para la masa y el radio. Determinados
  por  ecuaciones de estado mas realistas.
  \item Estructura interna de las Estrellas de Neutrones y sus consecuencias
  observacionales: estudio de los Sideramotos o disminución del
  período de rotación que pudiera deberse al choque entre la corteza
  y líquido superfluido.
   \item Origen, evolución y cotas de los campos magnéticos que ellas
   soportan.
  \item Determinación de las propiedades de la materia
  en regímenes críticos, alta densidad por encima de la densidad
nuclear.
  \item Existencia de ondas gravitacionales como consecuencia de los
  marcados efectos relativistas que traen consigo las ENs.
\end{itemize}

Al tratar de dar respuesta a estas preguntas se tropiezan con muchas
dificultades por lo que es preciso hacer modelos que describan de
forma aproximada los procesos que ocurren en el interior de las ENs.
Muchos de estos modelos se basan en tomar en cuenta diferentes
ecuaciones de estado y luego comparar los resultados con los que se
obtienen por medio de las observaciones.

%Los modelos propuestos hasta ahora difieren en varios aspectos
%siendo uno de los más importantes las ecuaciones de estado de la
%materia que compone la estrella pues para densidades superiores a
%las del núcleo atómico no se conoce bien el comportamiento de la
%materia. Varias ecuaciones de estado han sido estudiadas y se han
%obtenido diferentes resultados.

Como los campos magnéticos en las ENs son elevados resulta razonable
estudiar las ecuaciones de estado del gas de neutrones magnetizado.
En \cite{Aurora1} se ha estudiado el gas de neutrones relativista en
presencia de campo magnético interactuando con él a través del
momento magnético anómalo. Uno de los resultados obtenidos en este
estudio es que las presiones presentan anisotropías y el gas colapsa
en la dirección del campo.

%Al enfrentarnos al estudio de un objeto compacto en particular una
%ENs no podemos obviar los efectos que causa la materia en el
%espacio-tiempo por lo que debemos incluir la Teoría General de la
%Relatividad  para describir la evolución de la misma.

Debido a  que los sistemas con  anisotrop\'{\i}as, como el descrito
en \cite{Aurora1} en general no son estables en el universo un
objeto de este tipo debería colapsar o realizar una transici\'{o}n a
un estado estable.

Incluir la gravedad nos permite obtener información sobre el papel
que ella juega cuando aparecen estos tipo de anisotrop\'{i}as en las
presiones. Esta sería  la principal motivación de este trabajo y nos
llevaría a responder las siguientes preguntas:
\begin{itemize}
    \item ¿La Gravedad frenará o estimulará el colapso?
    %\item ¿Aparecerá o no el colapso magnético del gas de neutrones magnetizado?
    \item ¿Podrá este sistema evolucionar a una configuración estable?

\end{itemize}
Con el objetivo  de responder estas interrogantes, estudiaremos en
esta tesis,  el gas de neutrones  magnetizado autogravitante.
Utilizaremos una m\'{e}trica no estacionaria para obtener la
evoluci\'{o}n en el tiempo del sistema.

Por supuesto la formulaci\'{o}n de un modelo de una estrella
neutr\'{o}nica en presencia de campo magn\'{e}tico requiere del uso
de una m\'{e}trica mucho m\'{a}s complicada, que se traduce en un
problema n\'{u}merico m\'{a}s engorroso. Nuestro modelo aunque
simplificado puede dar informaci\'{o}n cualitativa muy interesante
sobre los procesos din\'{a}micos que le ocurren a un volumen local
en el interior de una ENs y constituye un primer paso en el estudio de un sistema tan complejo.\\

%Nuestro principal objetivo es hacer el estudio de la evolución de un
%volumen local para un gas de neutrones, para lo cual trabajaremos en
%un espacio Bianchi I caracterizado por una métrica tipo Kasner.

%El  objetivo de la tesis será por tanto estudiar la evolución
%dinámica de un gas relativista de neutrones en presencia de un campo
%magnético, tomando como punto de partida el estudio realizado en
%\cite{Aurora1}. En nuestro estudio incorporaremos los efectos de la
%RG, para lo cual trabajaremos en un espacio Bianchi I caracterizado
%por una métrica tipo Kasner que permite tomar en cuenta la
%anisotropía que introduce el campo.

En un trabajo anterior \cite{Alain2} fue estudiado el gas de
electrones magnetizado utilizando igualmente un espacio-tiempo
Bianchi I y una métrica Kasner. Este trabajo pretende entonces hacer
un estudio similar para el gas de neutrones.

La tesis tendrá 2 capítulos introductorios que consideramos básicos
para entender la problemática que nos ocupa. los capítulos
\textbf{3} y \textbf{4} se dedicaran a la parte original de la
tesis.

El contenido está distribuido de la siguiente forma:

\textbf{Capítulo 1} se describen las características generales de
las ENs así como su formación y evolución, además se discute el rol
de los campos magnéticos en estos sistemas y se discuten algunas
evidencias observacionales que existen actualmente sobre las ENs.

\textbf{Capítulo 2} se obtienen las ecuaciones de estado de un gas
degenerado de neutrones en presencia de un campo magnético
interactuando con él a través del momento magnético anómalo de los
neutrones.

\textbf{Capítulo 3} se utiliza la RG, introduciendo variables
dinámicas para obtener un sistema de ecuaciones diferenciales que
describen la evolución de un volumen local de una ENs.

\textbf{Capítulo 4} se estudian las soluciones del sistema de
ecuaciones en función de diferentes condiciones iniciales así como
las trayectorias en el espacio de fase. Posteriormente se exponen
las conclusiones.

%%%%%%%%%%%%%%%%%%%%%%%%%%%%%%%%%%%%%%%%%%%%%%%%%%%%%%%%%%%%%%%%%%%
% Capitulo 1
\chapter{Estrellas de Neutrones.}
\section{Características generales.}

El físico soviético L. D. Landau fue el primero en sugerir que
cuando la densidad de la materia es muy grande, los electrones se
ven obligados a fusionarse con los protones. Landau predijo que se
llegaría a una nueva configuración de equilibrio, en la que la
densidad de la materia es tan alta que los núcleos atómicos en
contacto forman un gigantesco núcleo.

En el año $1932$ el neutrón fue descubierto por Sir James Chadwick
lo cual aclaró el problema de la evolución estelar a altas
densidades. La existencia de un objeto estelar compuesto
fundamentalmente por neutrones fue propuesto de forma teórica por
primera vez en $1934$ por Walter Baade y Fritz Zwicky \cite{Zwicky}
en un artículo sobre la naturaleza de las supernovas, sugiriendo que
las ENs se formaban en este tipo de explosiones. Son de señalar los
trabajos de Oppenheimer y Volkoff realizados en 1939
\cite{Oppenheimer}, en los cuales estudian las posibles
configuraciones de equilibrio de una ENs utilizando un modelo de
estrella en el cual la presión del gas degenerado de neutrones
compensa la presión gravitacional como mismo ocurre con el gas de
electrones en las Enanas Blancas (EBs).

Actualmente se consideran como Estrellas de Neutrones (ENs)
\cite{Lattimer_1} a las estrellas con masas del orden de 1.5 veces
la masa del sol $(\mathrm{M}=1,5\,\mathrm{M_\odot})$, radios de
alrededor de $12\,\mathrm{km}$ ($R\sim12\,\mathrm{km}$) y densidades
comparables con las del núcleo atómico $(\rho\sim 10^{14}\,
\mathrm{g/cm^3})$. La temperatura superficial de una estrella de
neutrones se encuentra entre $3\times 10^5\, \kelvin$ y $10^6\,
\kelvin$ y la temperatura central es del orden de $10^8\,\kelvin$.
La energía de Fermi $(\mu_F)$ la podemos estimar conociendo que:
\[
\rho=\dfrac{Nm_n}{V} \Rightarrow
\dfrac{N}{V}=\dfrac{\rho}{m_n}=n\sim
\begin{cases}
6.022\times 10^{36}\,\metre\rpcubed, & \text{para $\rho_{sup}\sim10^{10}\,\kilogrampercubicmetre$},\\
4,21\times 10^{44}\, \metre\rpcubed, & \text{para
$\rho_{cent}\sim7\times 10^{17}\,\kilogrampercubicmetre$},
\end{cases}
\]
donde $\rho$ es la densidad, $N$ es el número de partículas, $m_n$
es la masa del neutrón y $V$ es el volumen. La energía de Fermi es
\cite{CRodriguez}:
\[
\mu_F=\dfrac{\hbar^2}{2m_n}(3\pi^2\dfrac{N}{V})^{(2/3)}\sim
\begin{cases}
1.06\times 10^{-16}\, \joule, & \text{en la superficie},\\
1,80\times 10^{-11}\, \joule, & \text{en el centro},
\end{cases}
\]
y la temperatura de Fermi $(T_F=\dfrac{\mu_F}{k_B})$ queda:
\[
T_F\sim\begin{cases}
7.69\times 10^{6}\, \kelvin, & \text{en la superficie},\\
1,30\times 10^{12}\, \kelvin, & \text{en el centro}.
\end{cases}
\]
Para que un gas fermiónico se pueda considerar degenerado tiene que
cumplirse la condición $\dfrac{T}{T_F}\ll1$, como vemos en la ENs
esto se cumple para regiones cercanas al centro donde
$\dfrac{T}{T_F}\sim10^{-4}$.

%Para una discusión sobre si las estrellas pueden considerarse como
%degeneradas ver Gravitacion pag 599 (624)

\begin{figure}[h!]
\begin{center}
\includegraphics[height=13cm,width=12cm,angle=0]{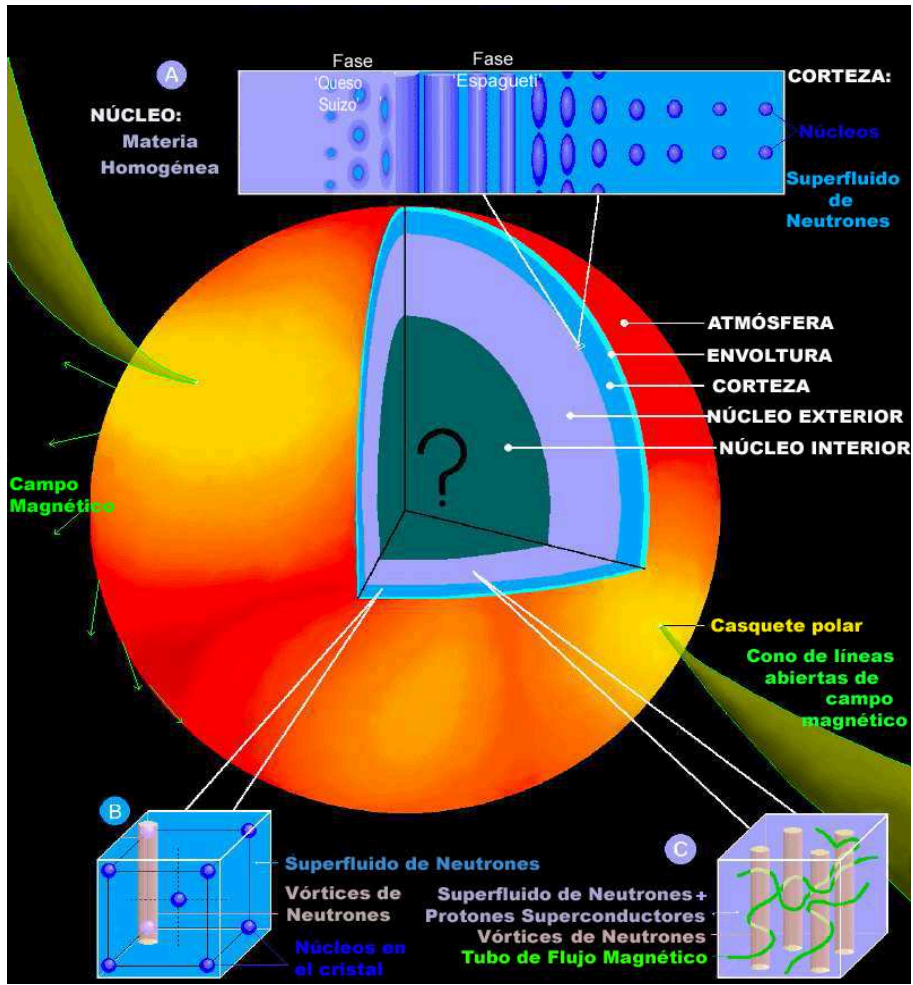}\\
\caption{\small{Regiones fundamentales en las que se divide una ENs
y posible composición.}}\label{ENs}
\end{center}
\end{figure}

Aunque la mayor parte de la estrella está formada por neutrones,
existen además protones y suficientes electrones para garantizar la
neutralidad de la estrella. Producto de las altas densidades también
se pueden encontrar toda una amplia gama de partículas en  su
interior. La estructura interna de una ENs consta de una serie de
capas de diferentes composiciones, la misma se divide en 5 partes
fundamentalmente \cite{Lattimer_1}. Una representación esquemática
se muestra en la figura (\ref{ENs}).

%\begin{enumerate}
%  \item Atmósfera.
%  \item Envoltura.
%  \item Corteza.
%  \item Núcleo exterior.
%  \item Núcleo interior.
%\end{enumerate}

La atmósfera y la envoltura contienen una cantidad despreciable de
masa, pero la atmósfera juega un papel importante en la forma del
espectro de los fotones que salen de la estrella, y la envoltura
tiene una influencia crucial en la liberación y el transporte de la
energía térmica de la superficie de la estrella. La corteza se
extiende entre 1 a $2\,\mathrm{km}$ por debajo de la superficie y
está formada fundamentalmente por núcleos atómicos. El núcleo de la
estrella contiene el $99\%$ de la masa de la estrella. El núcleo
exterior consiste en una combinación de nucleones, electrones y
muones. En el núcleo interior pueden existir partículas exóticas, se
piensa que incluso puede ser posible la transición a una fase mixta
de materia hadrónica y quarks libres.

\section{Formación y evolución.}

Cuando las estrellas llegan al final de su vida, producto del
agotamiento del combustible que mantiene las reacciones
termonucleares necesarias para garantizar el equilibrio, evolucionan
hacia un estado final dependiendo de su masa. Algunas explotan
dispersándose en el medio estelar, otras se contraen formando enanas
blancas, ENs o agujeros negros.

Las estrellas de neutrones surgen producto del colapso gravitatorio
del núcleo de una estrella masiva $(>8\,\mathrm{M}_{\odot})$. Al
final de su vida una estrella masiva expulsa la mayor parte de su
masa en una explosión conocida como supernova tipo II, es
precisamente en esta explosión donde surgen las ENs. En una estrella
ocurren reacciones de fusión en las cuales el hidrógeno se
transforma en helio, este por ser más pesado se va acumulando en el
núcleo, cuando éste alcanza una temperatura de $10^8\,\kelvin$ los 3
procesos alfa comienzan, formándose $^{12}_{6}\mathrm{C}$ mediante
las reacciones:
\begin{equation*}
^{4}_{2}\mathrm{He}+^{4}_{2}\mathrm{He} \  \rightarrow \
^{8}_{4}\mathrm{Be}+^{4}_{2}\mathrm{He} \  \rightarrow \
^{12}_{6}\mathrm{C} + \gamma + 7.367 \mathrm{MeV}.
\end{equation*}
Reacciones como estas se repiten formándose elementos más pesados
($^{20}\mathrm{O}$, $^{28}\mathrm{Si}$). Mientras más pesados sean
los elementos producidos en cada reacción menor es la energía que se
libera, las reacciones continúan hasta que se forma
$^{56}\mathrm{Fe}$ que es el núcleo con menor energía por nucleón.
La energía no puede continuar produciéndose por la fusión y el
núcleo de la estrella formado por hierro crece. Este aumento
continúa hasta sobrepasar el valor de $1,4\,\mathrm{M}_{\odot}$
(límite de Chandrasekhar), en este punto el núcleo comienza a
colapsar, pues la presión del gas degenerado de electrones no es
suficiente para contrarrestar la fuerza gravitatoria y las capas
superficiales son expulsadas hacia afuera, esta explosión es la
supernova.

En el núcleo de la estrella, protones y electrones son comprimidos
formándose neutrones $(n)$ y neutrinos $(\nu_e)$. Producto del
colapso la densidad aumenta y los núcleos atómicos se vuelven más
ricos en neutrones, los cuales comienzan a escapar de los núcleos.
Para densidades superiores $(\rho\sim10^{13}\,\mathrm{g/cm^3})$ los
núcleos remanentes se desintegran quedando la materia compuesta,en
su mayor parte, por neutrones. Los neutrones libres son inestables
con un tiempo de vida media cercano a los $15$ minutos decayendo a
través de la reacción $\beta^-: n \rightarrow p+e^-+\nu_e^-$.

En la materia a altas densidades el decaimiento $\beta^-$ no ocurre
gracias al Principio de Exclusión de Pauli \cite{Andreas}: Si todos
los niveles energéticos bajos de los electrones y protones están
ocupados, solo los neutrones suficientemente energéticos podrán
decaer.

Por otro lado, los protones y electrones energéticos podrán
combinarse mediante la reacción $\beta^+: e^-+p \rightarrow
n+\nu_e$. Entre los procesos $\beta^+$ y $\beta^-$ habrá un
equilibrio si los potenciales químicos de los neutrones, electrones
y protones cumplen la relación: $\mu_n=\mu_e+\mu_p$. El destino
final del núcleo de la estrella está determinado por su masa. Si
esta se mantiene por debajo de $3\,\mathrm{M}_{\odot}$ se habrá
formado una ENs, si es superior el núcleo continuará su colapso
formándose así un agujero negro.

\section{Campos magnéticos intensos.}
Las ENs son uno de los objetos astrofísicos a los cuales se les
atribuyen los mayores valores de campos magnéticos que existen en la
naturaleza. Estos
%Los campos magnéticos más intensos que existen en la naturaleza son
%los que se encuentran en las ENs,
superan por muchos órdenes de magnitud a los campos que son
familiares para los seres humanos. Por ejemplo, el campo terrestre
es $0,6\,\mathrm{G}$ en los polos, el campo permanente más intenso
que se ha obtenido en el laboratorio es de $10^5\,\mathrm{G}$
mientras que el campo de una ENs puede llegar hasta
$10^{18}\,\mathrm{G}$.

Existen varias hipótesis para explicar la formación de los campos
magnéticos en las ENs. El modo más simple de explicar su surgimiento
es producto de la compresión del flujo magnético durante el colapso
del núcleo de la estrella progenitora:
\begin{equation*}
\mathrm{\B}_{ENs}=\Bigl(\frac{\mathrm{R_0}}{\mathrm{R_{ENs}}}\Bigr)^2\mathrm{\B_0}
\sim (10^{11}-10^{13})\,\mathrm{G},
\end{equation*}
donde $\mathrm{\B}_{ENs}$ es el campo de la ENs, $\mathrm{\B_0}$ es
el campo del núcleo de la estrella progenitora,
$\mathrm{R_0}=(10^{5}-10^{6}) \, \mathrm{km}$ son los valores entre
los que se encuentran los radios del núcleo de la estrella
progenitora y $\mathrm{R_{ENs}}=10\,\mathrm{km}$ es el radio típico
de una ENs. Con este mecanismo se explican los campos magnéticos de
los radiopulsares, sin embargo hay observaciones que arrojan valores
mayores en los campos $(\sim10^{15}\,\mathrm{G})$ asociados a
objetos llamados Magnetars. Para explicar el surgimiento de tan
elevados campos se han propuesto mecanismos más complejos, como son
los procesos tipo dinamo que consisten básicamente en la generación
de corrientes internas producto de la rotación. La explicación y
determinación del origen, valor y evolución de los campos magnéticos
es un problema abierto para los astrofísicos y físicos de
partículas.

La cota máxima de los campos magnéticos la podemos estimar exigiendo
que la energía magnética sea menor que la energía gravitatoria:
\[
\frac{\B^2}{8\pi}\frac{4\pi
R^3}{3}<\frac{\mathrm{G_NM^2}}{\mathrm{R}} \ \ \Rightarrow \ \
\B\sim10^{18}\,\mathrm{G}.
\]
Estos campos magnéticos se encuentran en ENs jóvenes pues a medida
que la estrella envejece pierde energía rotacional y los campos
magnéticos disminuyen hasta valores de $10^9\,\mathrm{G}$.

\section{Evidencia observacional.}
En 1967 se confirma observacionalmente la existencia de las ENs
gracias a los esfuerzos de Joselyn Bell, estudiante de doctorado del
profesor Antony Hewish de Cambridge, que descubre la primera fuente
de radio (Pulsar), en el Radio Observatorio de dicha Universidad.
Este primer pulsar lleva el nombre de CP 9119. Actualmente se
conocen más de 1600 púlsares, que no son más que ENs que emiten
radiación electromagnética de forma periódica. El período mínimo de
una ENs lo podemos estimar igualando la fuerza centrífuga en el
ecuador con la fuerza gravitacional:
\[
m\omega^2\mathrm{R}=\frac{\mathrm{G_NMm}}{\mathrm{R}^2} \
\Rightarrow \
\mathrm{T}_{min}=2\pi\sqrt{\frac{\mathrm{R}^3}{\mathrm{G_NM}}}=0,5\,\milli\second.
\]
Este valor tan bajo del período es uno de los principales
indicadores de que estamos en presencia de una ENs. Los períodos de
los púlsares observados van desde $8,51\,\second$ (PSR J2144-3933
\cite{3933}) hasta $716\,\mathrm{Hz} \Rightarrow 1.3\,\milli\second$
(PSR J1748-2446ad) \cite{716} que es el pulsar más rápido que se
conoce. Recientemente ha sido descubierto una ENs con una frecuencia
de $1122\pm0.3\,\mathrm{Hz}$, la misma fue observada en el
transiente de rayos X XTE J1739-285  \cite{285}. Objetos compactos
que roten con períodos menores que $0,5\,\milli\second$ han hecho
pensar en la posibilidad de la existencia de estrellas de quarks
(EQs).

Las masas observadas para las ENs se encuentran entre
$1\,\mathrm{M_\odot}-2,1\,\mathrm{M_\odot}$ \cite{Lattimer_2} con un
valor promedio de $1.35\,\mathrm{M_\odot}$, así tenemos al pulsar
PSR J0751+1807 con una masa de $2,1\pm0,2\,\mathrm{M_\odot}$, el
caso mas típico se puede encontrar en el pulsar binario PSR 1913+16
\cite{Lattimer_2} en el cual las masas son
$(1.3867\pm0.0002)\,\mathrm{M_\odot}$ y
$(1.4414\pm0.0002)\,\mathrm{M_\odot}$ respectivamente. La masa más
pequeña es la del pulsar J1756-2251 \cite{2251} con un valor de
$(1.18\pm0.02)\,\mathrm{M_\odot}$.

Las mediciones de los radios de las ENs son muy imprecisas
$(9-15\,\mathrm{km})$ pues no se conoce exactamente la composición
química de la atmósfera ni la distancia exacta de las estrellas.

%Las mediciones de los radios y campos

%En la tabla \ref{tabla1} podemos ver los datos de algunos objetos.
%\begin{table}[!h]
%\begin{tabular}{|l||c|c|c|c|}\hline
%Objeto & Masa$[M_{\odot}]$ & Campo magnético$[\mathrm{G}]$ & Periodo[\second] & Radio$[\mathrm{km}]$ \\
%\hline\hline
%Tierra                &$3\times10^{-6}$& 0,6          & 86400 &  6371                          \\
%\hline
%Sol                   &      1         &  $1-2$    & $(2.16-3.1104)\times10^6$ & $6.95\times10^5$ \\
%\hline
%Sirio                 &                &              &       &                            \\
%\hline
%Nebulosa del Cangrejo &                &              &       &                            \\
%\hline
%\end{tabular}
%\caption{Datos de algunos objetos astrofísicos.}\label{tabla1}
%\end{table}

%%%%%%%%%%%%%%%%%%%%%%%%%%%%%%%%%%%%%%%%%%%%%%%%%%%%%%%%%%%%%%%%%%%
% Capitulo 2
\chapter{Gas de neutrones.}
\section{Ecuación de Dirac para partículas neutras con momento magnético anómalo.}
El gas de neutrones en presencia de un campo magnético ha sido
estudiado en \cite{Aurora1}, \cite{Guang}. En esta sección nos
limitaremos a exponer los resultados fundamentales que nos serán
útiles en la tesis.

Trabajaremos en el Ensemble Gran Canónico, considerando un
subsistema (en el contexto Astrofísico pudiera ser un elemento de
volumen dentro de la estrella), el cual se encuentra bajo la
influencia del campo magnético creado por el resto del sistema
$\vec{H}$. Producto de la presencia de este campo el subsistema se
polariza, creándose una magnetización en el medio (gas de neutrones)
que satisface la relación $\vec{H}=\vec{\B}-4\pi \vec{\M}$. El campo
$\vec{H}$ es externo al subsistema mientras que $\vec{\B}$ es
externo a cualquier partícula que se escoja dentro del subsistema,
pues esta siente en adición a $\vec{H}$, la contribución
$4\pi\vec{\M}$ del resto de las partículas del subsistema.

Para encontrar las ecuaciones de estado del gas de neutrones es
necesario calcular los autovalores de la energía de las partículas
que conforman el sistema. Para esto se resuelve la ecuación de Dirac
para partículas neutras con momento magnético anómalo:
%\reversemarginpar\marginpar{\tiny!!!Resolver esta ecuación
%para hallar los autovalores de la energía!!!!!}
\begin{equation}\label{Diraceq}
(\gamma_\mu\partial_\mu+m+iq\sigma_{\mu\lambda}F_{\mu\lambda})\psi=0,
\end{equation}
donde $\sigma_{\mu\lambda}=\frac{1}{2}(\gamma_\mu\gamma
_\lambda-\gamma _\lambda\gamma _\mu)$ es el tensor de spin y
$F_{\lambda\mu}$ es el tensor del campo electromagnético.

En la ecuación (\ref{Diraceq}) se ha tomado el convenio
$\hbar=\mathrm{c}=1$, en este sistema de unidades
$[\mathrm{L}]=[\mathrm{T}]=[\mathrm{M}^{-1}]$. A lo largo de la
tesis emplearemos este convenio en el manejo de las ecuaciones por
lo que en ninguna de estas aparecerán factores que contengan a
$\hbar$ o a $\mathrm{c}$. No haremos uso del convenio cuando estemos
estimando ordenes de magnitudes, expresando estas en el SI o en el
CGS lo cual será evidente por sus unidades.

De la solución de la ecuación (\ref{Diraceq}) se obtienen los
siguientes autovalores de la energía \cite{Guang}, \cite{Bagrov}:
\begin{equation}
E_n(p,\B,\eta)=\sqrt{p_3^2+(\sqrt{p_\bot^2+m_n^2}+\eta q\B)^2},
\end{equation}
donde $p_3, p_\bot$ son respectivamente las componentes del momentum
en las direcciones paralela y perpendicular al campo magnético
$(\B)$, $m_n$ es la masa del neutron, $q=-1.91\mu_N$ es el momento
magnético del neutron ($\mu_N=e/2m_p$ es el magneton nuclear),
$\eta=\pm1$ son los autovalores de $\sigma_3$ correspondientes a las
dos orientaciones del momento magnético con respecto al campo
magnético. No hemos tenido en cuenta las correcciones radiativas por
lo cual con\-si\-de\-ra\-re\-mos que el momento magnético se
mantiene constante para campos elevados.

El gran potencial termodinámico toma la forma
\begin{equation}
\Omega=-kT \ln\mathcal{Z},
\end{equation}
donde $k$ es la constante de Boltzman, $T$ es la temperatura,
$\mathcal{Z}=Tr(\hat{\rho})$ es la función de partición del sistema
y $\rho = e^{-(\hat{H}-\mu \hat{N})/kT}$, $\hat{H}$ es el
Hamiltoniano, $\mu$ el potencial químico y $\hat{N}$ el operador
número de partículas.

\section{Ecuaciones de estado.}
La forma del tensor energía-momentum de la materia en un campo
magnético externo constante es \cite{Alain}:
\begin{equation}\label{tensor_e-m}
\mathcal{T}^{i}_{\,\,\,\,k}=(T\frac{\partial{\Omega}}{\partial{T}}+\sum{\mu_{r}\frac{\partial{\Omega}}{
\partial{\mu_{r}}}})\delta^{i}_{\,\,4}\delta^{4}_{\,\,k}+4F^{il}F_{lk}\frac{\partial{\Omega}}{
\partial{F^{2}}}-\delta^{i}_{\,\,k}\Omega,
\end{equation}
Este tensor en el caso de limite de campo cero reproduce el tensor
$\mathcal{T}^{i}_{\,\,\,\,k}=P\delta^{i}_{\,\,k}-(P+U)\delta^{i}_{\,\,4}\delta^{4}_{\,\,k}$
de un fluido ideal.

Las componentes de este tensor son:
\begin{subequations}
\begin{eqnarray}
\mathcal{T}^{3}_{\,\,\,\,3}&=&P_3=-\Omega,
\\
\mathcal{T}^{1}_{\,\,\,\,1}&=&\mathcal{T}^{2}_{\,\,\,\,2}=-\Omega-\B\M,
\\
\mathcal{T}^{4}_{\,\,\,\,4}&=&-U=-TS-\mu N-\Omega,
\end{eqnarray}
\label{componentes de T}
\end{subequations}
donde $S$ es la entropía, $N=-\partial\Omega/\partial\mu$ es la
densidad de partículas, $\M=-\partial\Omega/\partial\B$ es la
magnetización, $U$ es la densidad de energía y $P_3$ es la presión
en la dirección del campo.

La expresión para el potencial termodinámico consta de dos términos:
\[
\Omega=\Omega_{sn}+\Omega_{Vn},
\]
el primero es la contribución estadística y el segundo la del vacío
\cite{Aurora1}, explícitamente tenemos:
\begin{equation}\label{omega_n}
\Omega_{sn}=-\frac{1}{4\pi^2\xi}\sum_{\eta=1,-1}^{}\int_0^\infty
p_\bot \mathrm{d}p_\bot \mathrm{d}p_3
\mathrm{ln}[f^+(\mu_n,\xi)f^-(\mu_n,\xi)],
\end{equation}
donde $\xi=1/k_BT$, $f^{\pm}(\mu_n,\xi)=(1+e^{(E_n\mp\mu_n)\xi})$
representan las contribuciones de las partículas y de las
antipartículas. La expresión para el término de vacío esta dada por
la expresión:
\begin{equation}\label{omega_v}
\Omega_{Vn}=\frac{1}{4\pi^2\xi}\sum_{\eta=1,-1}^{}\int_0^\infty
p_\bot \mathrm{d}p_\bot \mathrm{d}p_3 E_n,
\end{equation}
que es divergente pero se puede regularizar obteniéndose que para
$\B<10^{18}\,\mathrm{G}$ no es importante su contribución
\cite{Aurora1}, por lo que no se tendrá en cuenta en nuestro
estudio.

La expresión (\ref{omega_n}) se puede integrar fácilmente en nuestro
caso $(T=0)$ obteniéndose:
\begin{equation}\label{omega}
\Omega_{sn}=-\lambda\sum_{\eta=1,-1}^{}\biggl[\frac{\mu
f_\eta^3}{12}+ \frac{(1+\eta \beta)(5\eta \beta-3)\mu
f_\eta}{24}+\frac{(1+\eta \beta)^3(3-\eta
\beta)}{24}L_\eta-\frac{\eta \beta \mu^3}{6}s_\eta\biggr],
\end{equation}
donde hemos introducido las notaciones:
\[
f_\eta=\sqrt{\mu^2-(1+\eta\beta)^2}, \ \
s_\eta=\frac{\pi}{2}-\arcsin\biggr(\frac{1+\eta\beta}{\mu}\biggl), \
\ \mu=\frac{\mu_n}{m_n}
\]
\[
L_\eta=\ln\biggl(\frac{\mu+f_\eta}{1+\eta\beta}\biggr), \ \
\beta=\frac{\B}{\B_c},
\]
siendo $\B_c=m_n/q \simeq 1.56\times10^{20}\,\mathrm{G}$ el campo
crítico y
$\lambda=\dfrac{m_n^4}{4\pi^2\hbar^3\mathrm{c}^3}=4.11\times10^{36}
\mathrm{\,erg \,cm^{-3}}$.

A partir del Potencial Termodinámico $\Omega$ podemos obtener todas
las cantidades termodinámicas del sistema. En particular la densidad
de neutrones y la magnetización, de esta forma obtenemos que
$N=N_0\Gamma_N$ y $\M=\M_0\Gamma_M$ donde $N_0=\lambda/m_n$,
$\M_0=N_0q$ y los coeficientes $\Gamma_N,\Gamma_M$ vienen dados por:
\begin{subequations}
\begin{eqnarray*}
\Gamma_N&=& \sum_{\eta=1,-1}^{}\biggl[\frac{f_\eta^3}{3}+\frac{\eta
\beta(1+\eta\beta)f_\eta}{2}-\frac{\eta\beta \mu^2}{2}s_\eta\biggr],\\
\Gamma_M&=&-\sum_{\eta=1,-1}^{}\eta\biggl[\frac{(1-2\eta \beta)\mu
f_\eta}{6}-\frac{(1+\eta\beta)^2(1-\eta\beta/2)}{3}L_\eta+\frac{\mu^3}{6}s_\eta
\biggr].
\end{eqnarray*}
\end{subequations}

Teniendo en cuenta lo anterior, (\ref{componentes de T}) y
(\ref{omega}) podemos escribir las ecuaciones de estado para un gas
relativista degenerado de neutrones en un campo magnético externo en
la forma:
%\reversemarginpar\marginpar{\tiny!!!Debo obtener las EOS!!!!!}
\begin{subequations}
\begin{eqnarray}
U&=&\mu_n N+\Omega=\lambda \Gamma_U(\beta,\mu),
\\
p&=&-\Omega=\lambda \Gamma_P(\beta,\mu),
\\
M&=&\B \M=\lambda\beta \Gamma_M(\beta,\mu),
\end{eqnarray}
\label{EOS}
\end{subequations}
donde
\begin{subequations}
\begin{eqnarray*}
\Gamma_P&=&\sum_{\eta=1,-1}^{}\biggl[\frac{\mu f_\eta^3}{12}+
\frac{(1+\eta \beta)(5\eta \beta-3)\mu f_\eta}{24}+\frac{(1+\eta
\beta)^3(3-\eta
\beta)}{24}L_\eta-\frac{\eta \beta \mu^3}{6}s_\eta\biggr],\\
\Gamma_U&=&\mu \Gamma_N-\Gamma_P.
\end{eqnarray*}
\end{subequations}
%\reversemarginpar\marginpar{\tiny!!!!Hacer el tratamiento a partir
%de la física estadística de forma rigurosa!!!!!}

%%%%%%%%%%%%%%%%%%%%%%%%%%%%%%%%%%%%%%%%%%%%%%%%%%%%%%%%%%%%%%%%%%%
% Capitulo 3

\chapter{Relatividad General, métrica Kasner.}

\section{Relatividad General}

Debido a las altas densidades que se dan en las Estrellas de
Neutrones los efectos de la Relatividad General son muy marcados por
lo que se hace imprescindible introducir las ecuaciones de Einstein
para la descripción de este tipo de sistema. Una prueba de la acción
del campo gravitatorio en este sistema la podemos obtener fácilmente
estimando la velocidad de escape para una ENs típica
$(\mathrm{M=M_{\odot}}, \mathrm{R}=10\,\mathrm{km})$:
\[
 \mathrm{v_{esc}}=\Bigl(\frac{2 \mathrm{G_NM}}{ \mathrm{R}}\Bigr)^{1/2}\sim
 0.5\,\mathrm{c},
\]
donde $\mathrm{c}$ es la velocidad de la luz y $\mathrm{G_N}$ es la
constante gravitacional. Si la comparamos con la de la tierra
$(11.2\,\mathrm{km}\,\second^{-1})$ e incluso con la del sol
$(617.5\,\mathrm{km}\,\second^{-1})$ o la de una EB $(\sim
0.02\,\mathrm{c})$ podemos ver que la velocidad de escape de una ENs
es mucho mayor.

Asumiremos que nuestro sistema queda descrito por las ecuaciones de
Einstein las cuales vinculan el contenido de energía-materia del
sistema con la geometría del espacio-tiempo \cite{MTW}:
\begin{equation}
G^{\mu}_{\,\,\,\,\nu}=\kappa \mathcal{T}^{\mu}_{\,\,\,\,\nu},
\label{EE1}
\end{equation}
donde $\kappa=8\pi \mathrm{G_N}$ en el SI $\kappa$ tiene la
expresión $\kappa={\dfrac{8\pi \mathrm{G_N}}{c^4}}$,
$\mathrm{G_N}=6.67\times
10^{-11}\,\meter^{3}\kilogram^{-1}\second^{-2}$, $\mathrm{G}_{{\mu
\nu}}=R_{{\mu \nu}}-\frac{1}{2}R g_{{\mu \nu}}$ es el tensor de
Einstein el cual viene determinado por el tensor de Ricci $R_{{\mu
\nu}}$ y por el escalar de Ricci $R=R^{\mu}_{\,\,\,\,\mu}$, estos
últimos dependen de segundas derivadas de la métrica.
\begin{equation}
R_{\mu\nu}=\Gamma^{\alpha}_{\mu\nu,\alpha}-\Gamma^{\alpha}_{\mu\alpha,\nu}+\Gamma^{\alpha}_{\mu\nu}
\Gamma^{\beta}_{\alpha\beta}-\Gamma^{\beta}_{\mu\alpha}
\Gamma^{\alpha}_{\nu\beta}, \label{TF2}
\end{equation}
las cantidades $\Gamma^{\alpha}_{\mu\nu}$ son los índices de
Christoffel, que dependen de primeras derivadas de la métrica por la
fórmula,
\begin{equation}
\Gamma^{\alpha}_{\mu\nu}=\frac{g^{\alpha\beta}}{2}(g_{\beta\mu,\nu}+g_{\nu\beta
,\mu}-g_{\mu\nu,\beta}). \label{TF3}
\end{equation}
$\mathcal{T}^{\mu}_{\,\,\,\,\nu}$ es el tensor energía impulso (2.4)
que en nuestro caso lo podemos escribir como:
\begin{equation}
\mathcal{T}^{\mu}_{\,\,\,\,\nu}=(U+P)u^{\mu}u_{\nu}+P\delta^{\mu}_{\,\,\nu}+\Pi^{\mu}_{\,\,\nu},
\ \ P=p-\frac{2\B\M}{3}, \label{TE-M}
\end{equation}
en el cual
\begin{equation}
\Pi^{\mu}_{\,\,\nu}=diag[\Pi,\Pi,-2\Pi,0], \ \ \
\Pi=-\frac{\B\M}{3},\ \ \ \Pi^{\mu}_{\,\,\mu}=0. \label{TAn}
\end{equation}

Para hallar la forma del tensor de Einstein debemos tomar una
métrica apropiada, como ya dijimos es conveniente en el caso que nos
ocupa elegir una métrica tipo Kasner pues esta es compatible con la
anisotropía del gas de neutrones magnetizado:
%\marginpar{\tiny!!!!DEBO DAR UNA EXPLICACIÓN MAS DETALLADA DEL
%PORQUE ESCOJO LA MÉTRICA KASNER Y EL TENSOR E-M!!!!!}

\begin{equation}
{ds^2}=A(t)^2dx^{2}+B(t)^2dy^2+C(t)^2 d{z}^2-dt^2, \label{Metrica-K}
\end{equation}
con la métrica (\ref{Metrica-K}) y el tensor (\ref{TE-M}) se obtiene
que las componentes no nulas de la ecuación (\ref{EE1}) son:
\begin{subequations}
\begin{eqnarray}
-G^{x}_{\,\,x}&=&\frac{\dot{B}\dot{C}}{BC}+\frac{\ddot{B}}{B}+\frac{\ddot{C}}{C}=-\kappa(p-\B\M),
\\
-G^{y}_{\,\,y}&=&\frac{\dot{A}\dot{C}}{AC}+\frac{\ddot{A}}{A}+\frac{\ddot{C}}{C}=-\kappa(p-\B\M),
\\
-G^{z}_{\,\,z}&=&\frac{\dot{A}\dot{B}}{AB}+\frac{\ddot{A}}{A}+\frac{\ddot{B}}{B}=-\kappa
p,
\\
-G^{t}_{\,\,t}&=&\frac{\dot{A}\dot{B}}{AB}+\frac{\dot{A}\dot{C}}{AC}+\frac{\dot{B}\dot{C}}{BC}=\kappa
U.
\end{eqnarray}
\label{EE2}
\end{subequations}
La notación del punto significa derivada respecto el tiempo, por
ejemplo $\dot{A}=\dfrac{dA}{dt}$, $\ddot{A}=\dfrac{d^2A}{dt^2}$.

La ecuación de Einstein implica la conservación de la energía:
\begin{equation}\label{eq_cons}
\mathcal{T}^{\mu\nu}_{ \, \, \, \, \, \, \, \, \, ; \nu}=0.
\end{equation}
El punto y coma denota la derivada covariante, por ejemplo:
\begin{equation}
\mathcal{T}^\mu_{\,\,\,\,\nu ; \alpha}= \frac{\partial
\mathcal{T}^{\mu}_{\,\,\,\,\nu}}{\partial
x^{\alpha}}-\Gamma^{\gamma}_{\nu\alpha}\mathcal{T}^{\mu}_{\,\,\,\,\gamma}+\Gamma^{\mu}_{\gamma\alpha}\mathcal{T}^{\gamma
}_{\,\,\,\,\nu}.
\end{equation}
Tomando en cuenta la ecuación (\ref{eq_cons}) obtenemos:
\begin{equation}\label{eq_U[t]}
\dot{U}=\frac{\dot{C}(p+U)}{C}-(\frac{\dot{A}}{A}+\frac{\dot{B}}{B})(-\B\M+p+U).
\end{equation}
Tomemos además las ecuaciones de Maxwell:
\begin{equation}\label{Maxwell_eq}
F^{\mu\nu}\,_{;\nu}=0, \ \ F_{[\mu\nu;\alpha]}=0,
\end{equation}
así obtenemos:
\begin{equation}\label{Maxwell_eq_1}
\frac{\dot{A}}{A}+\frac{\dot{B}}{B}+\frac{1}{2}\frac{\dot{\B}}{\B}=0.
\end{equation}
Hemos obtenido las ecuaciones (\ref{EE2}, \ref{eq_U[t]},
\ref{Maxwell_eq_1}) en las que tenemos derivadas de primero y
segundo orden de las funciones $A, B, C$ y $U$. Para obtener un
sistema de ecuaciones diferenciales de primer orden cuya integración
numérica resulte mas sencilla introduciremos un conjunto de nuevas
variables.

\section{Variables dinámicas.}
Para la descripción dinámica de un elemento de volumen utilizaremos
las variables que usualmente se usan en cosmológica para
caracterizar la evolución del universo. Así tomaremos un sistema
comovil en el cual la cuatrivelocidad es $u^\alpha=\delta^\alpha _t$
y cumple que $u^\alpha u_\alpha =-1$.
%Introduciendo las notaciones
%\begin{equation}
%u_{(\mu;\nu)}=\frac{1}{2}(u_{\mu;\nu}+u_{\nu;\mu})  \ \ ; \ \
%u_{[\mu;\nu]}=\frac{1}{2}(u_{\mu;\nu}-u_{\nu;\mu})
%\end{equation}
%para la parte simétrica y antisimétrica de $u_{\mu;\nu}$.

%\reversemarginpar\marginpar{\tiny!!!!Explicar el significado de
%estas variables en cosmología así como su significado físico!!!!!}
Introduciremos las siguientes variables definidas a partir del
vector $u^{\alpha}$
\begin{equation}
h_{\alpha \beta}=g_{\alpha \beta}+u_{\alpha}u_{\beta},\label{AED1}
\end{equation}
$h_{\alpha \beta}$ se denomina tensor de proyección si este se
contrae con otro tensor, como resultado lo proyecta sobre un
3-espacio ortogonal a la 4-velocidad $u^\alpha$, cumpliéndose que
\begin{equation}
h_{\alpha}^{\nu}h_{\nu}^{\beta}=h_{\alpha}^{\beta}, \ \ \
h_{\alpha}^{\ \beta}u_{\beta}=0, \ \ \ h_{ \ \alpha}^{\alpha}=3.
\end{equation}
También se puede descomponer la derivada covariante
$u_{\alpha;\beta}$ en las siguientes partes irreducibles,
\begin{equation}
u_{\alpha;\beta}=\sigma_{\alpha \beta} +\omega_{\alpha \beta
}+\frac{1}{3}\Theta h_{\alpha \beta}-a_{\alpha}u_{\beta},
\label{AED3}
\end{equation}
donde $\sigma_{\alpha \beta}$ es un tensor simétrico de traza nula,
$\omega_{\alpha \beta}$ un tensor antisimétrico, y
$u^{\alpha}\sigma_{\alpha \beta}=u^{\alpha}\omega_{\alpha \beta}=0$,
siendo,
\begin{subequations}
\begin{eqnarray}
\Theta&=&u^{\alpha}_{;\alpha}\, , \ \ \ \\
a_{\alpha}&=&u_{\alpha;\beta}u^{\beta}, \ \ \ \\
\sigma_{\alpha \beta}&=& \frac{1}{2}(u_{\alpha; \mu}h^{\mu}_{\
\beta}+u_{\beta; \mu}
h^{\mu}_{\ \alpha})-\frac{1}{3} \Theta h_{\alpha \beta}, \ \ \ \\
\omega_{\alpha \beta}&=& \frac{1}{2}(u_{\alpha ; \mu}h^{\mu}_{\
\beta}-u_{\beta; \mu}h^{\mu}_{\ \alpha}).
\end{eqnarray}\label{AED4}
\end{subequations}
La cantidad $\Theta$ se llama escalar de expansión, la misma
caracteriza la velocidad a la cual el elemento de volumen ortogonal
a $u^\mu$ se expande o contrae, $a_{a}$ es la 4-aceleración del
fluido. El tensor $\sigma_{ab}$ es el rango de deformación, el cual
describe la manera en la que el elemento de volumen ortogonal a la
cuatrivelocidad $(u^\mu)$ cambia su forma, la magnitud de este
tensor viene dada por:
\begin{equation}
\sigma^{2}=\frac{1}{2}\sigma_{\alpha \beta}\sigma^{\alpha \beta},
\label{AED5}
\end{equation}
y $\omega_{\alpha \beta}$ es el tensor de vorticidad, el cual es una
medida de la rotación presente en la materia y es conveniente
definirlo vectorialmente como:
\begin{equation}
\omega^{\alpha}=\frac{1}{2}\eta^{\alpha \beta \gamma \delta
}u_{\beta}\omega_{\gamma \delta}, \label{AED6}
\end{equation}
aquí $\eta$ es el tensor totalmente antisimétrico, además se
satisface que $u^{\alpha}\omega_{\alpha}=0$, y la magnitud de la
vorticidad viene dada por,
\begin{equation}
\omega^{2}=\omega^{\alpha}\omega_{\alpha}=\frac{1}{2}\omega^{\alpha
\beta }\omega_{\alpha \beta}, \label{AED7}
\end{equation}
si la vorticidad es cero entonces se dice que el vector del campo
$u^{\alpha}$ es i\-rro\-ta\-cio\-nal. Esto es algunas veces
conveniente para combinar la magnitud del rango de la deformación
con la expansión, quedando el tensor de expansión como,
\begin{equation}
\Theta_{\alpha \beta}=\sigma_{\alpha \beta}+\frac{1}{3}\Theta
h_{\alpha \beta}. \label{AED8}
\end{equation}
En el caso de la métrica (\ref{Metrica-K}) y la 4-velocidad
$u^\alpha =\delta^\alpha _t$, la 4-aceleración se anula, el escalar
de expansión $\Theta$ y el tensor de deformación $\sigma^\mu_\nu$
toman la forma:
\begin{equation}\label{ec_teta_K}
\Theta=\frac{\dot{A}}{A}+\frac{\dot{B}}{B}+\frac{\dot{C}}{C} \ \ , \
\ \sigma^\mu_\nu=diag[\sigma^x_x,\sigma^y_y,\sigma^z_z,0],
\end{equation}
donde
\begin{subequations}
\begin{eqnarray}
\sigma^x_x&=&\frac{2\dot{A}}{3A}-\frac{\dot{B}}{3B}-\frac{\dot{C}}{3C},\\
\sigma^y_y&=&\frac{2\dot{B}}{3B}-\frac{\dot{A}}{3A}-\frac{\dot{C}}{3C},\\
\sigma^z_z&=&\frac{2\dot{C}}{3C}-\frac{\dot{A}}{3A}-\frac{\dot{B}}{3B}.
\end{eqnarray}
\label{sigma_comp}
\end{subequations}
Haremos el siguiente cambio de notación para mayor comodidad en el
tratamiento de las ecuaciones:
\begin{equation}
\sigma^x_x\rightarrow\Sigma_1,  \ \ \sigma^y_y\rightarrow\Sigma_2, \
\ \sigma^z_z\rightarrow\Sigma_3.
\end{equation}
Haciendo uso de estas variables dinámicas y teniendo en cuenta las
ecuaciones (\ref{eq_U[t]}) y (\ref{Maxwell_eq_1}) obtenemos el
siguiente sistema de ecuaciones:

\begin{subequations}
\begin{eqnarray}
\dot{U}&=&-(U+p-\frac{2}{3}{\B}{\M})\Theta-\B\M \Sigma_3,
\\
\dot{\Sigma_2}&=&-\frac{\kappa \B\M}{3}-\Theta\Sigma_2,
\\
\dot{\Sigma_3}&=&\frac{2}{3}\kappa \B\M- \Theta\Sigma_3,
\\
\dot{\Theta}&=&\kappa (\B\M+\frac{3}{2}(U-p))-\Theta^2,
\\
\dot{\beta}&=&\frac{2}{3}\beta (3\Sigma_3-2\Theta).
\end{eqnarray}
\label{EE3}
\end{subequations}
Además de (3.7d) se obtiene que
\begin{equation}\label{constrain}
-\Sigma_2^2-\Sigma_2\Sigma_3+\frac{\Theta^2}{3}-\Sigma_3^2=\kappa U.
\end{equation}

Notemos que hemos obtenido un sistema de ecuaciones diferenciales no
lineales de primer orden en las variables $U, \beta, \Theta,
\Sigma_2,\Sigma_3$ más la ligadura (\ref{constrain}). La solución de
este sistema describe la evolución dinámica de un volumen local de
un gas de neutrones magnetizado que podría ser un volumen local en
el interior de una ENs.

%%%%%%%%%%%%%%%%%%%%%%%%%%%%%%%%%%%%%%%%%%%%%%%%%%%%%%%%%%%%%%%%%%%
% Capitulo 4
\chapter{Ecuaciones dinámicas.}

\section{Variables adimensionales.}

%The governing equations of the most commonly studied cosmological
%models are a system of autonomous ordinary differential equations
%(ODEs). Since our main goal is to give a qualitative description of
%these models, a dynamical systems approach is undertaken. Usually, a
%dimensionless (logarithmic) time variable, ô , is introduced so that
%the models are valid for all times (i.e., ô assumes all real
%values). A normalised set of variables are then chosen for a number
%of reasons. First, this normally leads to a compact dynamical
%system. Second, these variables are well-behaved and often have a
%direct physical interpretation. Third, due to a symmetry in the
%equations, one of the equations decouple (in general relativity the
%expansion is used to normalize the variables in ever expanding
%models whence the Raychaudhuri equation decouples) and the resulting
%simplified reduced system is then studied. The singular points of
%the reduced system then correspond to dynamically evolving
%self-similar cosmological models. More precisely, using the
%dimensionless time variable and a normalised set of variables, the
%governing ODEs define a flow and the evolution of the cosmological
%models can then be analysed by studying the orbits of this flow in
%the physical state space, which is a subset of Euclidean space. When
%the state space is compact, each orbit will have a non-empty á-limit
%set and ù-limit set, and hence there will be a both a past attractor
%and a future attractor in the state space.

En este capítulo estudiaremos la dinámica del gas de neutrones
magnetizado que equivale al volumen local de una ENs. Nuestro
interés fundamental es responder la pregunta inicial de si en el
marco de la Relatividad General el colapso obtenido en
\cite{Aurora1} se mantiene.

Para realizar la descripción dinámica del volumen local de la
estrella de neutrones introduciremos un conjunto de variables
adimensionales con las cuales obtendremos un sistema de ecuaciones
más compacto y con un significado físico mas claro, de esta manera
introduzcamos las nuevas variables:
% \reversemarginpar\marginpar{\tiny!!!Dar una explicación del
%significado que tiene el tiempo adimensional !!!!!}
\begin{equation}\label{def_tau}
H=\frac{\Theta}{3},  \ \ \frac{d}{d\tau}=\frac{1}{H_0}\frac{d}{dt},
\end{equation}
y las nuevas funciones adimensionales:
\begin{equation}\label{adim_var}
\HH=\frac{H}{H_0}, \ \ S_2=\frac{\Sigma_2}{H_0}, \ \
S_3=\frac{\Sigma_3}{H_0}, \ \ \beta=\frac{\B}{\B_c},
\end{equation}
donde $H_0$ es una constante que por conveniencia la escogeremos
como $3H_0^2=\kappa\lambda \Rightarrow
|H_0|=1.66\times10^{-4}\,\mathrm{cm^{-1}}$, notemos que en
cosmología $H_0=0.59\times10^{-28}\mathrm{cm^{-1}}$ es la constante
de Hubble cuyo inverso es una medida de la escala del universo
$(1/H_0=1.69\times10^{28}\,\mathrm{cm})$, en nuestro caso tenemos
que $1/H_0\sim 6\,\mathrm{km}$, magnitud razonable para nuestro
sistema. $S_2$ y $S_3$ están relacionados con la segunda y tercera
componente del tensor $\sigma_{\alpha \beta}$. El nuevo tiempo
adimensional $(\tau)$ lo definimos a partir de la ecuación
(\ref{def_tau}), notemos que este tiempo puede tomar valores tanto
positivos como negativos, dependiendo su signo del de
$H_0=\pm\sqrt{\dfrac{\kappa\lambda}{3}}$, para una mayor
profundización en el significado del signo del tiempo adimensional
ver el A.

Comprobar que las variables antes definidas no tienen dimensión es
muy simple si recordamos que los coeficientes de la métrica
(\ref{Metrica-K}) $(A, B, C)$ son adimensionales, además de
(\ref{ec_teta_K}) y (\ref{sigma_comp}) notamos que las variables
$\Theta, \Sigma_2$ y $\Sigma_3$ tienen unidades de inverso de
longitud al igual que $H_0$.

Si sustituimos las variables (\ref{adim_var}) en el sistema de
ecuaciones (\ref{EE3}) ob\-te\-ne\-mos:
\begin{subequations}
\begin{eqnarray}
\mu_{,\tau}&=&\frac{1}{\Gamma_{U,\mu}}\biggl[(2\HH-S_3)(\Gamma_M-2\Gamma_{U,\beta})\beta-3\HH(\Gamma_P+\Gamma_U)
\biggr],
\\
S_{2,\tau}&=&-\beta \Gamma_M-3S_2\HH,
\\
S_{3,\tau}&=&2\beta \Gamma_M-3S_3\HH,
\\
\HH_{,\tau}&=&\beta
\Gamma_M-\frac{3\Gamma_P}{2}-\frac{1}{2}S_2S_3-\frac{3}{2}\HH^2-\frac{1}{2}(S_2^2+S_3^2),
\\
\beta_{,\tau}&=&2\beta(S_3-2\HH),
\end{eqnarray}
\label{SED}
\end{subequations}
la coma indica derivada respecto al tiempo adimensional $(\tau)$,
por ejemplo $S_{3,\tau}=\dfrac{dS_3}{d\tau}$. Hemos cambiado la
variable $U$ por $\mu$, por (\ref{EOS}) sabemos que $U=U(\beta, \mu)
\Rightarrow
U_{,\tau}=\lambda(\Gamma_{U,\mu}\mu_{,\tau}+\Gamma_{U,\beta}\beta_{,\tau})$
de donde podemos despejar $\mu_{,\tau}$.

Además de (\ref{constrain}) tiene que cumplirse que:
\begin{equation}\label{vinculo}
-S_2^2-S_3^2-S_2S_3+3\HH^2=3\Gamma_U.
\end{equation}
El sistema de ecuaciones diferenciales (\ref{SED}) lo resolveremos
numéricamente empleando el programa Maple. Sus soluciones nos darán
el comportamiento en el tiempo del elemento de volumen del gas
magnetizado de neutrones. De (\ref{vinculo}) podemos despejar $\HH$
obteniendo dos raíces, de las cuales escogeremos la negativa que
garantiza (ver \ref{vol_local}) la condición de colapso.

\section{Soluciones numéricas y discusión física}
\subsection{Comportamiento de las funciones.}
Las soluciones que buscamos son las que nos dan el colapso del gas
magnetizado de neutrones que podría implicar el colapso de un
volumen de la ENs. Para garantizar la condición de colapso basta con
exigir en las ecuaciones(\ref{SED}) que la expansión inicial sea
negativa $(\HH(0)<0)$ pues su signo es el que nos indica si el
volumen esta colapsando o expendiéndose. De (\ref{adim_var}) y
(\ref{ec_teta_K}) se obtiene que el promedio del volumen local
$(V=ABC)$ lo podemos expresar como:
\begin{equation}\label{vol_local}
V=V(0)\exp(3\int_{\tau_0}^\tau\HH d\tau).
\end{equation}
Según el comportamiento de los coeficientes de la métrica podemos
clasificar las singularidades como \cite{dynamical}:
\begin{enumerate}
  \item tipo punto ($A,B,C\rightarrow0$),
  \item tipo cigarro (dos de los coeficientes métricos tienden a cero y el tercero se va a
  infinito),
  \item tipo barril (dos de los coeficientes métricos tienden a cero y el tercero tiende a un valor
  finito),
  \item tipo pancake (uno de los coeficientes métricos tienden a cero y los otros dos tienden a un valor
  finito).
\end{enumerate}
Estas denominaciones se refieren al cambio de forma del elemento de
volumen. Fueron introducidas por Thorne (1967) en estudios sobre
modelos Bianchi I, pero pueden aplicarse a casos generales.

Para investigar el tipo de colapso a partir del comportamiento de
los coeficientes espaciales de la métrica observemos que estos y las
variables dinámicas antes definidas ($\HH+S_i, i=1,2,3$)  están
relacionados a través de las siguientes expresiones:
\begin{subequations}
\begin{eqnarray}
A(\tau)&=A_0\exp[\int(S_1+\HH)d\tau],\\
B(\tau)&=B_0\exp[\int(S_2+\HH)d\tau],\\
C(\tau)&=C_0\exp[\int(S_3+\HH)d\tau],
\end{eqnarray}
\end{subequations}
donde $A_0, B_0 ,C_0$ son constantes, estas relaciones se obtienen
de (\ref{ec_teta_K}) y (\ref{sigma_comp}).

Para la solución del sistema (\ref{SED}) utilizaremos diferentes
condiciones iniciales típicas para las ENs, por ejemplo:
$\mu=2\Rightarrow \rho\sim10^{15}\,\mathrm{g/cm^3} $,
$\beta_0=10^{-2}-10^{-5}$ para campos de entre $10^{18}\mathrm{G}$ y
$10^{15}\mathrm{G}$. Impondremos siempre la condición de colapso del
volumen $\HH(0)<0$ y tomaremos $S_2(0)=0,\pm1$, $S_3(0)=0,\pm1$
co\-rres\-pon\-dien\-tes a casos con deformación inicial cero y
deformación inicial en la dirección de los ejes $y$ o $z$.

\begin{figure}[h!]
\begin{center}
\includegraphics[height=7.5cm,width=10cm,angle=0]{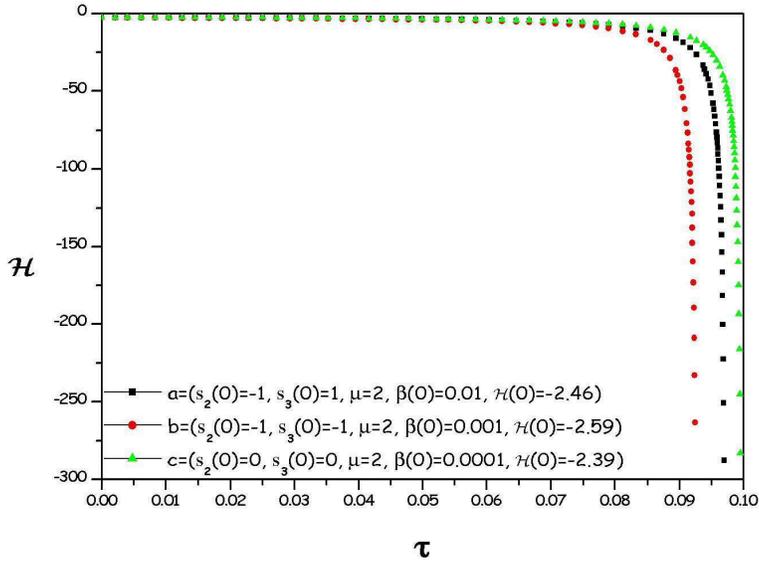}
\caption{\footnotesize{Comportamiento de $\HH$ vs $\tau$ para
diferentes condiciones iniciales.}}\label{H}
\end{center}
\end{figure}

\begin{figure}[h!]
\begin{center}
\includegraphics[height=7.5cm,width=10cm,angle=0]{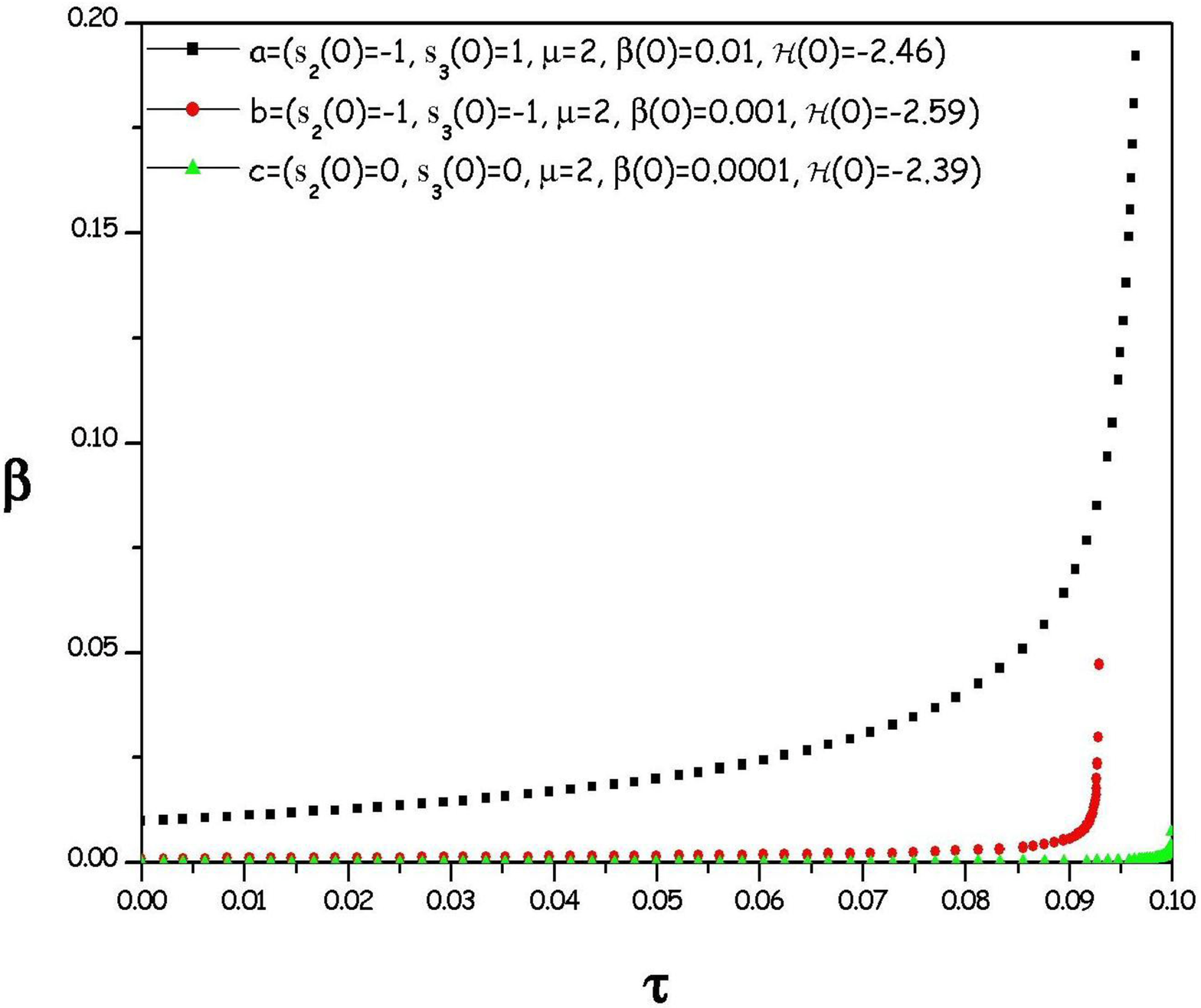}
\caption{\footnotesize{El comportamiento del campo magnético
$(\beta=\B/\B_c$) muestra una tendencia a incrementarse,
manteniéndose el valor del campo por debajo del campo
crítico.}}\label{b}
\end{center}
\end{figure}

La solución de $\HH$(figura \ref{H}) para distintas condiciones
iniciales nos muestra que $\HH\rightarrow-\infty$ independientemente
de las condiciones iniciales. El campo magnético tiende a aumentar
manteniéndose por debajo del campo crítico. Su comportamiento para
las condiciones iniciales estudiadas se puede apreciar en la figura
\ref{b}.

%\begin{figure}[!ht]
%\begin{minipage}[t]{.6\textwidth}
%\includegraphics[height=7.5cm,width=7.5cm,angle=0]{grf_h}
%\caption{\footnotesize{Comportamiento de $(\HH)$ vs $\tau$ para
%diferentes condiciones iniciales. Notese que la mayoría de los
%gráficos coinciden}} \label{H}
%\end{minipage}
%\hfil
%\begin{minipage}[t]{.6\textwidth}
%\includegraphics[height=7.5cm,width=7.5cm,angle=0]{grf_b}
%\caption{\footnotesize{El comportamiento del campo magnético
%$(\beta$ vs $\tau$) muestra una tendencia a incrementarse,
%manteniéndose el valor del campo por debajo del del campo crítico
%$(\beta<1)$.}}\label{b}
%\end{minipage}
%\end{figure}

\begin{figure}[!ht]
\begin{minipage}[t]{.6\textwidth}
\includegraphics[height=7cm,width=7cm,angle=0]{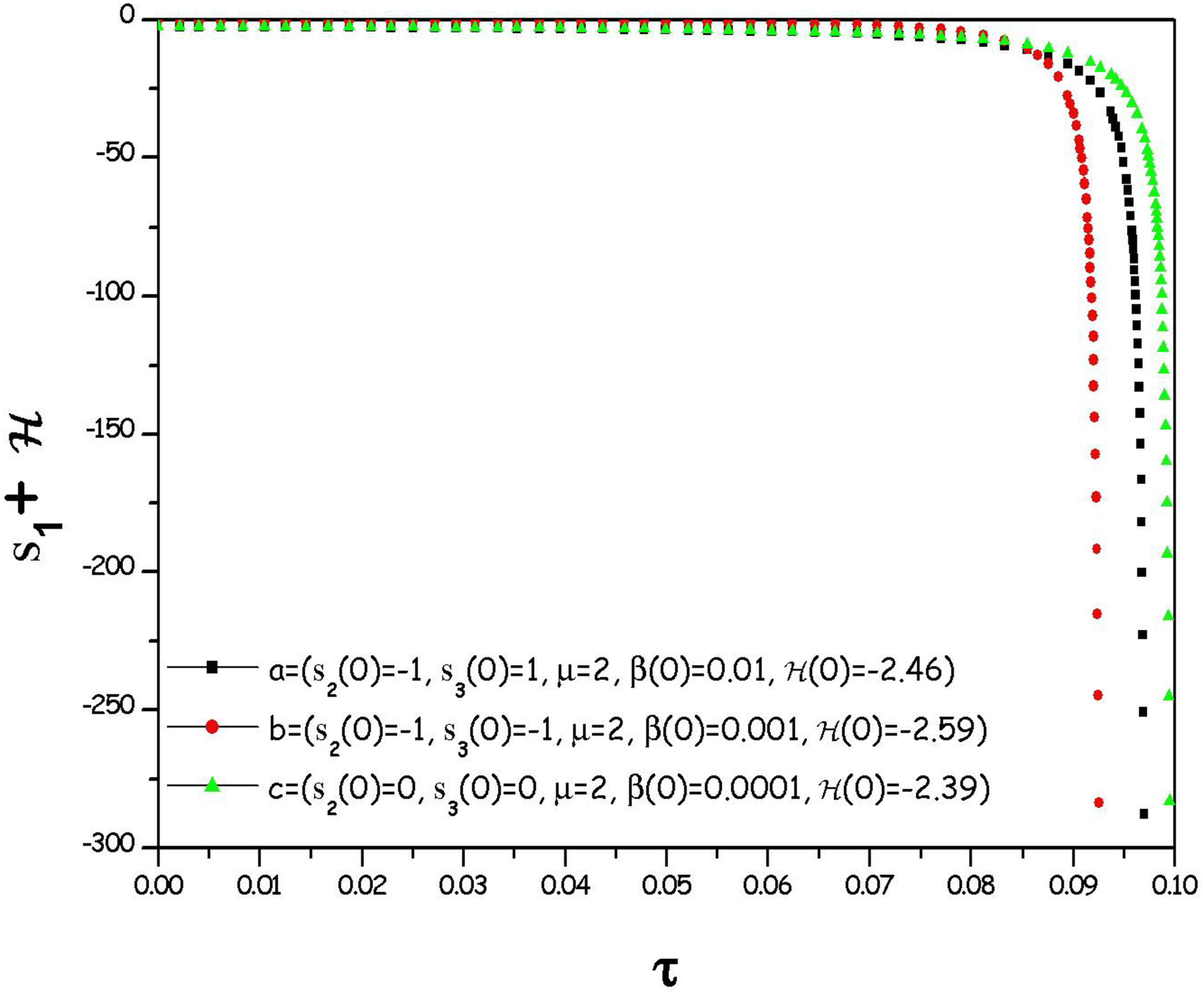}
%\caption{\footnotesize{Gráfico de $(S_1+\HH)$ vs $\tau$.}}
%\label{S1}
\end{minipage}
\hfil
\begin{minipage}[t]{.6\textwidth}
\includegraphics[height=7cm,width=7cm,angle=0]{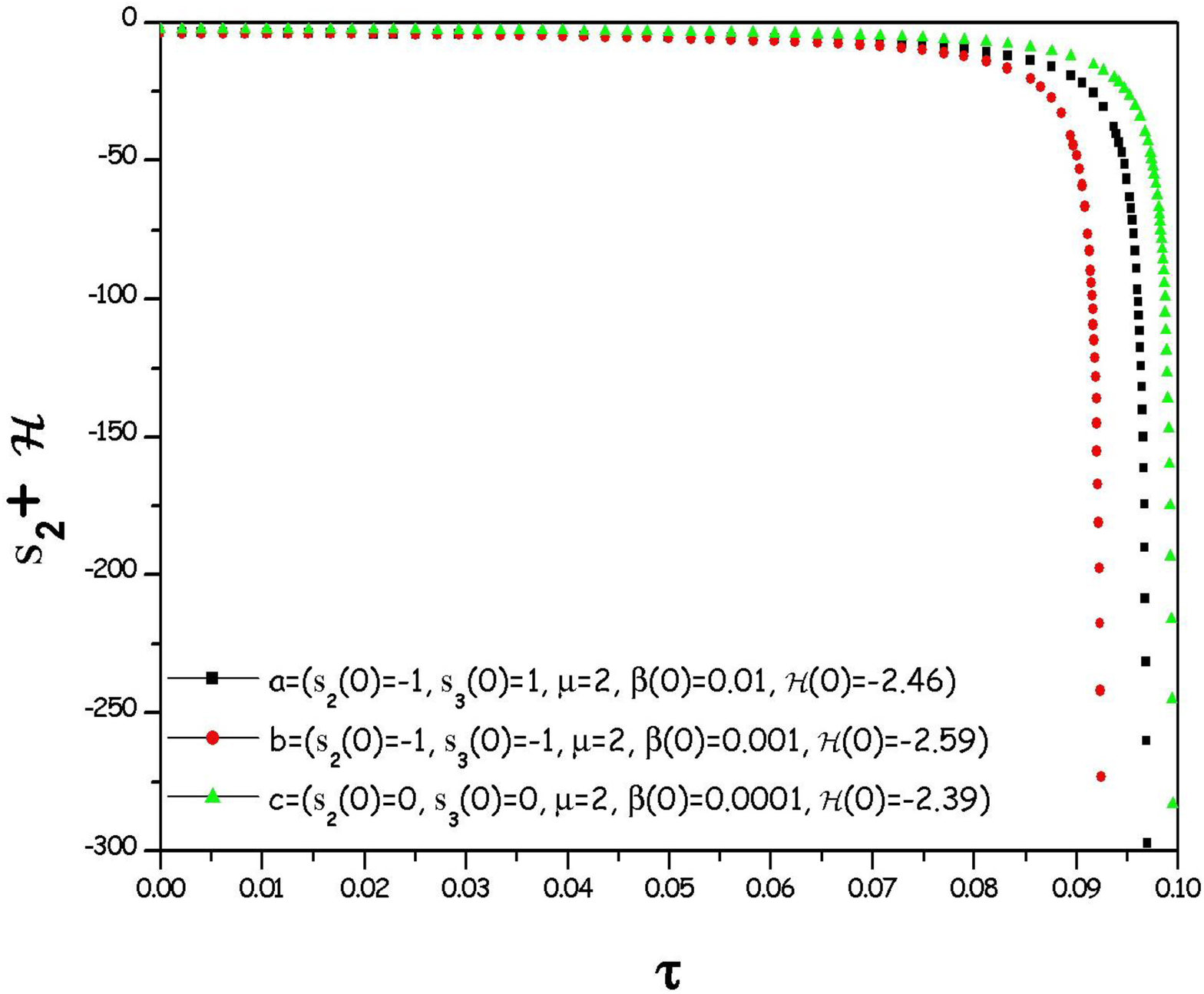}
%\caption{\footnotesize{Gráfico de $(S_2+\HH)$ vs $\tau$.}}
%\label{S2}
\end{minipage}
\caption{\footnotesize{Comportamiento de $(S_1+\HH)$ y de
$(S_2+\HH)$ vs $\tau$. Podemos observar la tendencia a $-\infty$
además de distintos tiempos de colapso para diferentes condiciones
iniciales}} \label{S1S2}
\end{figure}

\begin{figure}[!ht]
\begin{center}
\includegraphics[height=7.5cm,width=10cm,angle=0]{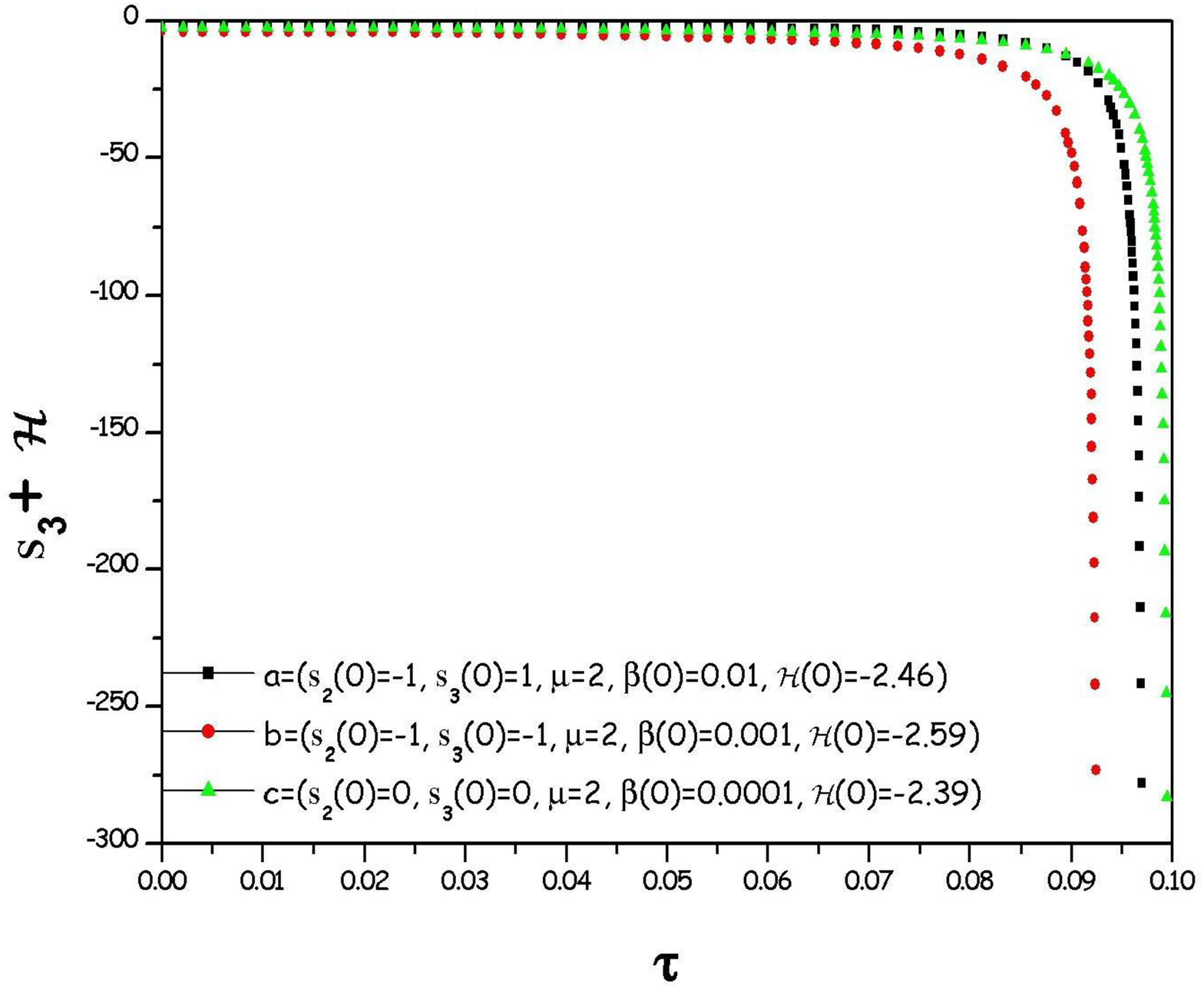}
\caption{\footnotesize{Gráfico de $(S_3+\HH)$ vs $\tau$. Igualmente
so observa que $(S_3+\HH) \rightarrow -\infty$}}\label{S3}
\end{center}
\end{figure}

De las soluciones del sistema de ecuaciones (\ref{SED}) obtenemos
los gráficos (\ref{S1S2}), (\ref{S3}) cuyo comportamiento nos da la
forma en que colapsa el elemento de volumen.

Podemos observar como los $S_i+\HH$ se van a $-\infty$ por lo que
los coeficientes espaciales de la métrica tienden a cero ($A, B,
C\rightarrow0$) de donde podemos inferir que el volumen colapsa en
una singularidad tipo punto.

\subsection{Espacio de fase.}
Si utilizamos la ecuación (\ref{vinculo}) podemos reducir el sistema
(\ref{SED}) a un sistema de ecuaciones en las variables $S_3, \beta,
\mu, \HH$ quedando:
\begin{subequations}
\begin{eqnarray}
\mu_{,\tau}&=&\frac{1}{\Gamma_{U,\mu}}\biggl[(2\HH-S_3)(\Gamma_M-2\Gamma_{U,\beta})\beta-3\HH(\Gamma_P+\Gamma_U)
\biggr],
\\
S_{3,\tau}&=&2\beta \Gamma_M-3S_3\HH,
\\
\HH{,\tau}&=&\beta \Gamma_M+\frac{3}{2}(\Gamma_U-\Gamma_P)-3\HH^2,
\\
\beta_{,\tau}&=&2\beta(S_3-2\HH),
\end{eqnarray}
\label{SED_2}
\end{subequations}
notemos que la única ecuación que ha cambiado es la  de $\HH$.

\begin{figure}[h!]
\begin{center}
\includegraphics[height=14cm,width=14cm,angle=0]{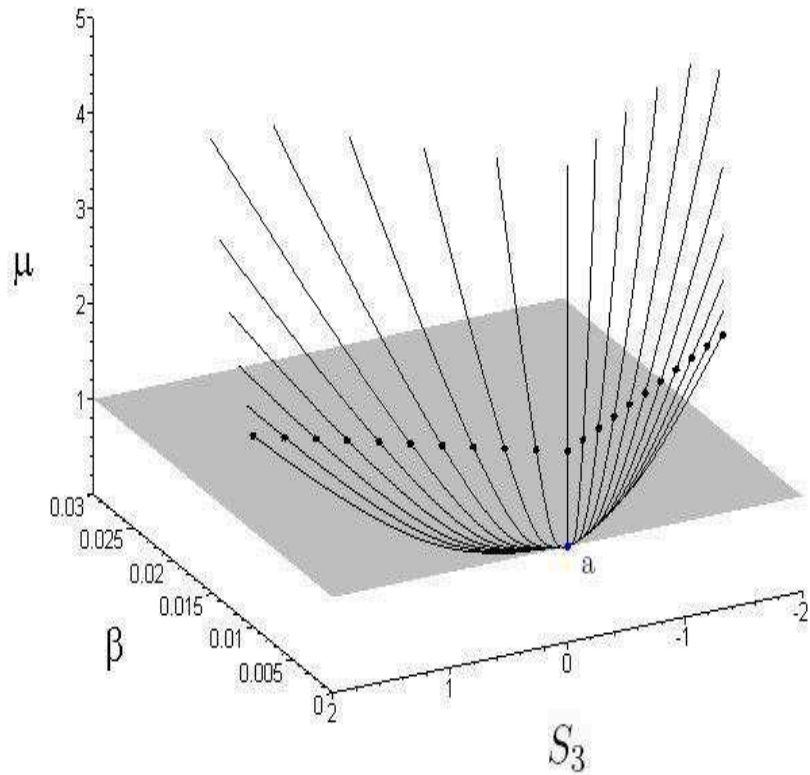}\\
\caption{\small{Trayectorias en la sección del espacio de fase
$(S_3, \beta, \mu)$. Los puntos sin marca representan las
condiciones iniciales, el punto \textquotedblleft
a\textquotedblright\  representa el atractor .}}\label{esp_fase}
\end{center}
\end{figure}

Las trayectorias en la sección $(S_3, \beta, \mu)$  del espacio de
fase se muestran en la figura (\ref{esp_fase}). La evolución del
sistema queda determinada por el signo de $H_0$. Para $\tau<0
\Rightarrow H_0=-\sqrt{\kappa\lambda/3}$ el sistema evoluciona hacia
el atractor estable (punto marcado con la a), mientras que para
$\tau>0 \Rightarrow H_0=\sqrt{\kappa\lambda/3}$ la evolución es
hacia una singularidad. Un estudio similar se realizó para las demás
secciones del espacio de fase obteniéndose que las coordenadas del
atractor son: $(S_3=0, \beta=0, \mu=1, \HH=0)$.

\section{Conclusiones.}
Hemos estudiado la evolución de un volumen local de un gas
magnetizado de neutrones bajo condiciones que pueden ser las que
existan en el interior de una ENs. Para aproximarnos a la solución
de este problema hemos utilizado las ecuaciones de estado de dicho
gas y las ecuaciones de Einstein-Maxwell junto a la ecuación de
conservación de la energía, obteniendo de esta forma un sistema de
ecuaciones diferenciales no lineales. El sistema obtenido lo hemos
puesto en función de nuevas variables que permiten reducir el orden
de las derivadas a la vez que dan una mejor interpretación de los
resultados. Para la solución del sistema de ecuaciones diferenciales
hemos introducido nuevas variables adimensionales, realizando los
cálculos numéricos con el programa Maple.

Los principales resultados obtenidos son:
\begin{enumerate}
\item La dinámica del volumen magnetizado de neutrones presenta una
solución singular que equivale a un colapso de la materia. Este
resultado se obtiene para todas las condiciones iniciales
estudiadas. Con ello res\-pon\-de\-mos la pregunta de que si los
efectos de la RG al ser tomados en cuenta favorecerían o no el
colapso magnético descrito en \cite{Aurora1}. Como podemos ver la
gravitación hace que el colapso se manifieste y ocurra en forma de
punto.
\item A diferencia del estudio dinámico realizado para el gas de
electrones magnetizado \cite{Alain2} el tipo de singularidad que
aparece en este caso es de tipo punto y no cigarro. Este resultado
sugiere que los efectos de la gravedad para un gas de neutrones
magnetizado son más marcados que para el gas de electrones,
cambiando la forma de la singularidad.
\item El espacio de fase muestra que el sistema evoluciona, para $H_0<0$, a un punto de
equilibrio, es decir una configuración estable.
\item El campo magnético tiende a aumentar, manteniendo valores por debajo de campo crítico.
\end{enumerate}

\subsection{Direcciones del trabajo futuro.}
Resultaría interesante estudiar un volume de gas de neutrones,
electrones y protones que obedezcan el equilibrio $\beta$ y la
neutralidad de carga, lo que describiría de manera mas realista las
condiciones en el interior de una ENs.

Otros espacio-tiempo como Bianchi V, VII ó IX permitirían tomar en
cuenta más grados de libertad, acercándose más a la descripción de
lo que le ocurre a un volumen local de una ENs.

%%%%%%%%%%%%%%%%%%%%%%%%%%%%%%%%%%%%%%%%%%%%%%%%%%%%%%%%%%%%%%%%%%%
% Conclusiones
%\input{Conclusiones.tex}
%%%%%%%%%%%%%%%%%%%%%%%%%%%%%%%%%%%%%%%%%%%%%%%%%%%%%%%%%%%%%%%%%%%
%\newpage
%\thispagestyle{empty} \mbox{}
%\newpage
% Conclusiones
%\input{Conclus.tex}
%%%%%%%%%%%%%%%%%%%%%%%%%%%%%%%%%%%%%%%%%%%%%%%%%%%%%%%%%%%%%%%%%%%
%\newpage
%\thispagestyle{empty} \mbox{}
%\newpage
\def\thesection{\Alph{chapter}.\arabic{section}}
\appendix
% Anexo 1
\chapter{Significado del tiempo adimensional $(\tau)$.}

El tiempo adimensional $(\tau)$ se define a partir de la ecuación
(\ref{def_tau}):
\begin{equation}\label{definicion de tau}
\dfrac{d}{d\tau}=\dfrac{1}{H_0}\dfrac{d}{dt} \ \ \Rightarrow \ \
\tau=H_0t,
\end{equation}
en la cual podemos notar que el signo de $\tau$ depende del signo de
$t$ y del signo de $H_0$. Para comprender el sentido físico del
tiempo adimensional $\tau$ observemos que de (\ref{adim_var}) y
(\ref{ec_teta_K}) obtenemos:
\begin{subequations}
\begin{eqnarray}
\HH
H_0=&H&=\frac{1}{3}(\frac{\dot{A}}{A}+\frac{\dot{B}}{B}+\frac{\dot{C}}{C})=\frac{1}{3}\frac{d}{dt}\ln(V),\\
&V&=ABC\Rightarrow \boxed{\frac{V}{V_0}=e^{3\int\limits_0^tHdt}},
\end{eqnarray}
\label{relacion de HH, V ,H0}
\end{subequations}
y despejando $\HH$ de (\ref{vinculo}) tenemos:
\begin{equation}\label{ec de H}
\HH=\pm\frac{1}{3}\sqrt{3\Gamma_U+S_2^2+S_3^2+S_2S_3}, \ \ \
\end{equation}

Si fijamos $t\geqslant0$, de (\ref{relacion de HH, V ,H0}) tenemos
la siguiente interpretación física:
\[
I=3\int\limits_0^tHdt\Rightarrow
\begin{cases}
I>0\Rightarrow V>V_0\Rightarrow  \text{expansión},\\
I<0\Rightarrow V<V_0\Rightarrow  \text{colapso}
\end{cases}
\]

\vspace{1cm}

Las posibles combinaciones de signo se dan en el cuadro A,
%(\ref{tabla1}),
en el cual aparecen resaltados los convenios que
hemos escogido en esta tesis.
\begin{table}[!ht]
\begin{center}
%\label{tabla1}\caption{Posibles combinaciones de signos entre $\HH$,
%$H_0$ y $\tau$ para $t>0$.}
\begin{tabular}{|c|c|c|c|}\hline
Casos & Expansión & Volumen elemental & $\tau$\\ \hline

$H>0$ & $\HH>0$ y $H_0>0$  & Expansión & $\tau>0$
\\ \cline{2-2} \cline{4-4}

& \textcolor[rgb]{1.00,0.00,0.00}{$\HH<0$ y $H_0<0$} & $V>V_0$&\textcolor[rgb]{1.00,0.00,0.00}{$\tau<0$} \\
\hline

$H<0$ &\textcolor[rgb]{1.00,0.00,0.00}{$\HH<0$ y $H_0>0$}  & Colapso
&\textcolor[rgb]{1.00,0.00,0.00}{ $\tau>0$}
\\ \cline{2-2} \cline{4-4}

& $\HH>0$ y $H_0<0$ & $V<V_0 $&  $\tau<0$ \\ \hline

$H=0$ & $\HH=0$ & $V=V_0$ &  \\

\hline
\end{tabular}
\end{center}
\end{table}

Tabla A: Posibles combinaciones de signos entre $\HH$,
$H_0$ y $\tau$ para $t>0$.
Notemos que el sistema comienza a evolucionar a partir de
$t=\tau=0$, para $\tau>0$ el sistema colapsa y para $\tau<0$ el
sistema tiende hacia un estado diluido como se puede ver en el
gráfico (\ref{esp_fase}).

%\lstinputlisting{Star.f} %%%%esto es por si quiero poner el codigo que esta en el archivo Star.f

%%%%%%%%%%%%%%%%%%%%%%%%%%%%%%%%%%%%%%%%%%%%%%%%%%%%%%%%%%%%%%%%%%%
% Anexo 2
%\input{Anexo2.tex}
%%%%%%%%%%%%%%%%%%%%%%%%%%%%%%%%%%%%%%%%%%%%%%%%%%%%%%%%%%%%%%%%%%%
% Referencias
%%%%%%%%%%%%%%%%%%%%%%%%%%%%%%%%%%%%%%%%%%%%%%%%%%%%%%%%
%                       Bibliografía                   %
%%%%%%%%%%%%%%%%%%%%%%%%%%%%%%%%%%%%%%%%%%%%%%%%%%%%%%%%

%%%%%%%%%%%%%%%%%%%%%%%%%%%%%%%%%%%%%%%%%%%%%%%%%%%%%%%%%%%%%%%%%%%

\begin{thebibliography}{500}

\bibitem{MTW} Charles W. Misner, KipS. Thorne, John. Archibald Wheeler, {\it Gravitation},
ed W.H.Freeman and Company, NY (1973). %p 618.


%\cite{Martinez:2003dz}
\bibitem{Aurora1}
  A.~P.~Martinez, H.~P.~Rojas and H.~J.~Mosquera Cuesta,
  %``Magnetic collapse of a neutron gas: Can magnetars indeed be formed,''
  Eur.\ Phys.\ J.\  C {\bf 29}, 111 (2003)
  [arXiv:astro-ph/0303213].
  %%CITATION = EPHJA,C29,111;%%

%\bibitem{Aurora1} A.  Pérez Martinez, H. Pérez  Rojas and H.J. Mosquera
%Cuesta 2003 {\it\ Eur. \ Phys.\ J.\ }, {\bf C 29},   111-123.

%\cite{Ulacia Rey:2007kc}
\bibitem{Alain2}
 A.~Ulacia Rey, A.~Perez Martinez and R.~A.~Sussman,
  %``Local dynamics and gravitational collapse of a self-gravitating magnetized
  %Fermi gas,''
  Gen.\ Rel.\ Grav.\  {\bf 40} (2008) 1499
  [arXiv:0708.0593 [gr-qc]].
  %%CITATION = GRGVA,40,1499;%%



%\bibitem{Alain2} A. Ulacia Rey, A. Pérez Martínez and Roberto A.
%Sussman 2007 Local dynamics and gravitational collapse of a
%self-gravitating magnetized Fermi gas {\it Preprint} gr-qc/0708.0593
%v2.

\bibitem{Zwicky} W.~Baade and F.~Zwicky,  Proc. Nat. Acad. Sci.,
{\bf 20}, 259 (1934).

\bibitem{Oppenheimer} J.~R.~Oppenheimer and G.~M.~Volkoff,  Phys. Rev. {\bf 55}, 374 (1939).

%\cite{Lattimer:2004pg}
\bibitem{Lattimer_1}
  J.~M.~Lattimer and M.~Prakash,
  %``The physics of neutron stars,''
  Science {\bf 304}, 536 (2004)
  [arXiv:astro-ph/0405262].
  %%CITATION = SCIEA,304,536;%%


%\bibitem{Lattimer_1} J.M. Lattimer and M. Prakash 2004  The Physics
%of Neutron Stars, {\it Preprint} astro-ph/0405262 v1.

\bibitem{CRodriguez} Carlos Rodriguez Castellanos, María Teresa
Pérez Maldonado, {\it Introduccíon a la física estadística}, ed
Felix Varela, La Habana (2002).

%\cite{Reisenegger:2008et}
\bibitem{Andreas}
  A.~Reisenegger,
  %``Neutron stars and their magnetic fields,''
  arXiv:0802.2227 [astro-ph].
  %%CITATION = ARXIV:0802.2227;%%

%\bibitem{Andreas}  2008, Neutron stars and their magnetic
%fields {\it Preprint} astro-ph/0802.2227v1.

%\cite{Gil:2000ke}
\bibitem{3933}
  J.~Gil and D.~Mitra,
  %``Vacuum gaps in pulsars and PSR J2144-3933,''
  arXiv:astro-ph/0010603.
  %%CITATION = ASTRO-PH/0010603;%%

%\bibitem{3933} Janusz Gil and Dipanjan Mitra 2000 Vacuum gaps in pulsars and PSR
%J2144-3933, {\it Preprint} astro-ph/0010603v1.

%\cite{Hessels:2006ze}
\bibitem{716}
  J.~W.~T.~Hessels, S.~M.~Ransom, I.~H.~Stairs, P.~C.~C.~Freire, V.~M.~Kaspi and F.~Camilo,
  %``A Radio Pulsar Spinning at 716 Hz,''
  arXiv:astro-ph/0601337.
  %%CITATION = ASTRO-PH/0601337;%%


%\bibitem{716} Jason W. T. Hessels, Scott M. Ransom, Ingrid H. Stairs,Paulo C. C. Freire,
%Victoria M. Kaspi, and Fernando Camilo 2006 A Radio Pulsar Spinning
%at $716\mathrm{Hz}$ {\it Preprint} astro-ph/0601337v1.

%\cite{Kaaret:2006gr}
\bibitem{285}
  P.~Kaaret {\it et al.},
  %``Discovery of 1122 Hz X-Ray Burst Oscillations from the Neutron-Star X-Ray
  %Transient XTE J1739-285,''
  arXiv:astro-ph/0611716.
  %%CITATION = ASTRO-PH/0611716;%%

%\bibitem{285} P. Kaaret, Z. Prieskorn, J.J.M. in't Zand, S. Brandt, N. Lund,
%S. Mereghtti, D. Götz, E. Kuulkers and J.A. Tomsick 2007 Evidence
%for 1122 HZ X-Ray Burst oscillations from the Neytron-Star X-Ray
%Trannsient XTE J1739-285 {\it Preprint} astro-ph/0611716v2.

%\cite{Lattimer:2006xb}
\bibitem{Lattimer_2}
  J.~M.~Lattimer and M.~Prakash,
  %``Neutron Star Observations: Prognosis for Equation of State Constraints,''
  Phys.\ Rept.\  {\bf 442}, 109 (2007)
  [arXiv:astro-ph/0612440].
  %%CITATION = PRPLC,442,109;%%


%\bibitem{Lattimer_2} James M. Lattimer and Madappa Prakash Neutron Star
%Observations: Prognosis for Equation of State Constraints 2006 {\it
%Preprint} astro-ph/0612440v1.

%\cite{Faulkner:2004ha}
\bibitem{2251}
  A.~J.~Faulkner {\it et al.},
  %``PSR J1756-2251: a new relativistic double neutron star system,''
  Astrophys.\ J.\  {\bf 618} (2004) L119
  [arXiv:astro-ph/0411796].
  %%CITATION = ASJOA,618,L119;%%

%\bibitem{2251} A. J.Faulkner, M. Kramer, A. G. Lyne, R. N. Manchester, M. A.
%McLaughlin, I. H. Stairs, G. Hobbs, A. Possenti, D. R. Lorimer, N.
%D'Amico,, F. Camilo and M. Burgay 2004 PSR J1756-2251: a new
%relativistic double neutron star system {\it Prerprint}
%astro-ph/0411796v1.

\bibitem{Guang} Guang-Jun Mao, Akira Iwamoto and Zhu-Xia Li, Chin. J. Astron.
Astrophys. {\bf 3}, No. 4, 359-374 (2003).

\bibitem{Bagrov} V.~G.~Bagrov, D.~M.~Gitman, {\it Exact solutions of
relativistic wave equations} (Kluwer Academic Publ (1990).

\bibitem{Alain} Alain Ulacia Rey, {\it Dinámica de una fuente magnetizada y
autogravitante} Tesis presentada en opción al grado de Master en
Ciencias Físicas, Universidad de la Habana (2006).


\bibitem{dynamical} J.~Wainwright, G.~F.~R.~Ellis, {\it Dynamical Systems in
Cosmology}, ed Cambridge University Press (1997).




\end{thebibliography}
\end{document}